\documentclass[12pt,preprint]{aastex}
\shortauthors{Krawczynski et al.}
\shorttitle{Multiwavelength Observations of 1ES~1959+650}
\begin{document}
\title{Multiwavelength Observations of Strong Flares From the TeV-Blazar
1ES~1959+650}
\author{
H.~Krawczynski\altaffilmark{1}, 
S.~B.~Hughes\altaffilmark{1}, 
D.~Horan\altaffilmark{2}, 
F.~Aharonian\altaffilmark{3}, 
M.~F.~Aller\altaffilmark{4},
H.~Aller\altaffilmark{4}, 
P.~Boltwood\altaffilmark{5}, 
J.~Buckley\altaffilmark{1}, 
P.~Coppi\altaffilmark{6}, 
G.~Fossati\altaffilmark{7},
N.~G\"otting\altaffilmark{8}
J.~Holder\altaffilmark{9}, 
D.~Horns\altaffilmark{3}, 
O.~M.~Kurtanidze\altaffilmark{10,11}
A.~P.~Marscher\altaffilmark{12},
M.~Nikolashvili\altaffilmark{10},
R.~A.~Remillard\altaffilmark{13},
A.~Sadun\altaffilmark{14}, 
M.~Schr\"oder\altaffilmark{2}
}
\altaffiltext{1}{Washington University in St. Louis, Physics Department, 
1 Brookings Drive CB 1105,St. Louis, MO 63130}
\altaffiltext{2}{F.\ Lawrence Whipple Observatory, Harvard-Smithsonian CfA, Amado, AZ 85645}
\altaffiltext{3}{Max Planck Institut f\"ur Kernphysik, Postfach 103980, D-69029 Heidelberg, Germany}
\altaffiltext{4}{University of Michigan, Department of Astronomy,
Ann Arbor, MI, 48109-1090}
\altaffiltext{5}{1655 Main Street, Stittsville, Ontario, Canada K2S 1N6}
\altaffiltext{6}{Yale University, P.O. Box 208101, New Haven, CT 06520-8101, USA}
\altaffiltext{7}{Rice University, MS 108, 6100 Main street, Houston, TX 77005}
\altaffiltext{8}{Universit\"at Hamburg, Institut f\"ur
Experimentalphysik, Luruper Chaussee 149,
D-22761 Hamburg, Germany}
\altaffiltext{9}{University of Leeds, Department of Physics, Leeds, LS2 9JT,
Yorkshire, UK}
\altaffiltext{10}{Abastumani Astrophysical Observatory, 383762, Abastumani, Republic of Georgia}
\altaffiltext{11}{Landessternwarte Heidelberg-K\"onigstuhl and 
Astrophysikalisches Institut Postdam}
\altaffiltext{12}{Institute for Astrophysical Research, Boston University, 725 Commonwealth Avenue, Boston, MA 02215-1401}
\altaffiltext{13}{Center for Space Research, Massachusetts Institute of Technology, Cambridge, MA 02139}
\altaffiltext{14}{Department of Physics, University of Colorado at Denver, Denver, CO 80217}
\begin{abstract}
Following the detection of strong TeV $\gamma$-ray flares from the BL Lac object
1ES~1959+650 with the Whipple 10~m Cherenkov telescope on May 16 and 17, 2002, we performed intensive 
Target of Opportunity (ToO) radio, optical, X-ray and TeV $\gamma$-ray observations 
from May 18, 2002 to August 14, 2002.
Observations with the X-ray telescope {\it RXTE} {\it (Rossi X-ray Timing Explorer)}  and the 
Whipple and HEGRA {\it (High Energy Gamma Ray Astronomy)} $\gamma$-ray telescopes 
revealed several strong flares, enabling us to sensitively test the X-ray/$\gamma$-ray  
flux correlation properties. Although the X-ray and $\gamma$-ray fluxes seemed to be 
correlated in general, we found an ``orphan'' $\gamma$-ray flare that
was not accompanied by an X-ray flare. 
While we detected optical flux variability with the Boltwood and Abastumani observatories, 
the data did not give evidence for a correlation between the optical flux variability with 
the observed X-ray and $\gamma$-ray flares. Within statistical errors of about 0.03 Jy at 14.5 GHz 
and 0.05~Jy at 4.8 GHz, the radio fluxes measured with the University of Michigan Radio 
Astrophysical Observatory (UMRAO) stayed constant throughout the campaign;
the mean values agreed well with the values measured on May 7 and June 7, 
2002 at 4.9 GHz and 15 GHz with the 
Very Large Array (VLA), and, at 4.8 GHz with archival flux measurements.
After describing in detail the radio, optical, X-ray and $\gamma$-ray
light curves and Spectral Energy Distributions (SEDs) we present
initial modeling of the SED with a simple Synchrotron Self-Compton (SSC) 
model. 
With the addition of another TeV blazar with good broadband data,
we consider the set of all TeV blazars to begin to look for a connection
of the jet properties to the properties of the central accreting black hole
thought to drive the jet. Remarkably, the temporal and spectral
X-ray and $\gamma$-ray emission characteristics of TeV blazars 
are very similar, even though the masses estimates of their central black 
holes differ by up to one order of magnitude. 
\end{abstract}
\keywords{galaxies: BL Lacertae objects: individual (1ES~1959+650) --- 
galaxies: jets --- gamma rays: observations}
\section{Introduction}
\label{intro}
The EGRET {\it (Energetic Gamma Ray Experiment Telescope)} 
detector on board of the {\it Compton Gamma-Ray Observatory} 
discovered 100~MeV--$\sim$1~GeV $\gamma$-ray emission from 66 blazars, 
mainly from Flat Spectrum Radio Quasars and Unidentified Flat 
Spectrum Radio Sources  \cite{Hart:99}.
Ground-based Cherenkov telescopes discovered TeV $\gamma$-ray emission from
6 blazars, 4 of which are not EGRET sources.
The electromagnetic emission of these Active Galactic Nuclei (AGNs)
is dominated by a non-thermal continuum with a low-energy synchrotron component
and a high-energy Inverse Compton component 
(see Coppi (1999), Sikora \& Madejski (2001), Krawczynski (2003a) for recent reviews).
The TeV sources all belong to the class of BL Lac objects, 
blazars with relatively low luminosity but with Spectral 
Energy Distributions (SEDs) that peak at extremely high energies.

\begin{figure}[bh]
\begin{center}
\resizebox{12cm}{!}{
\plotone{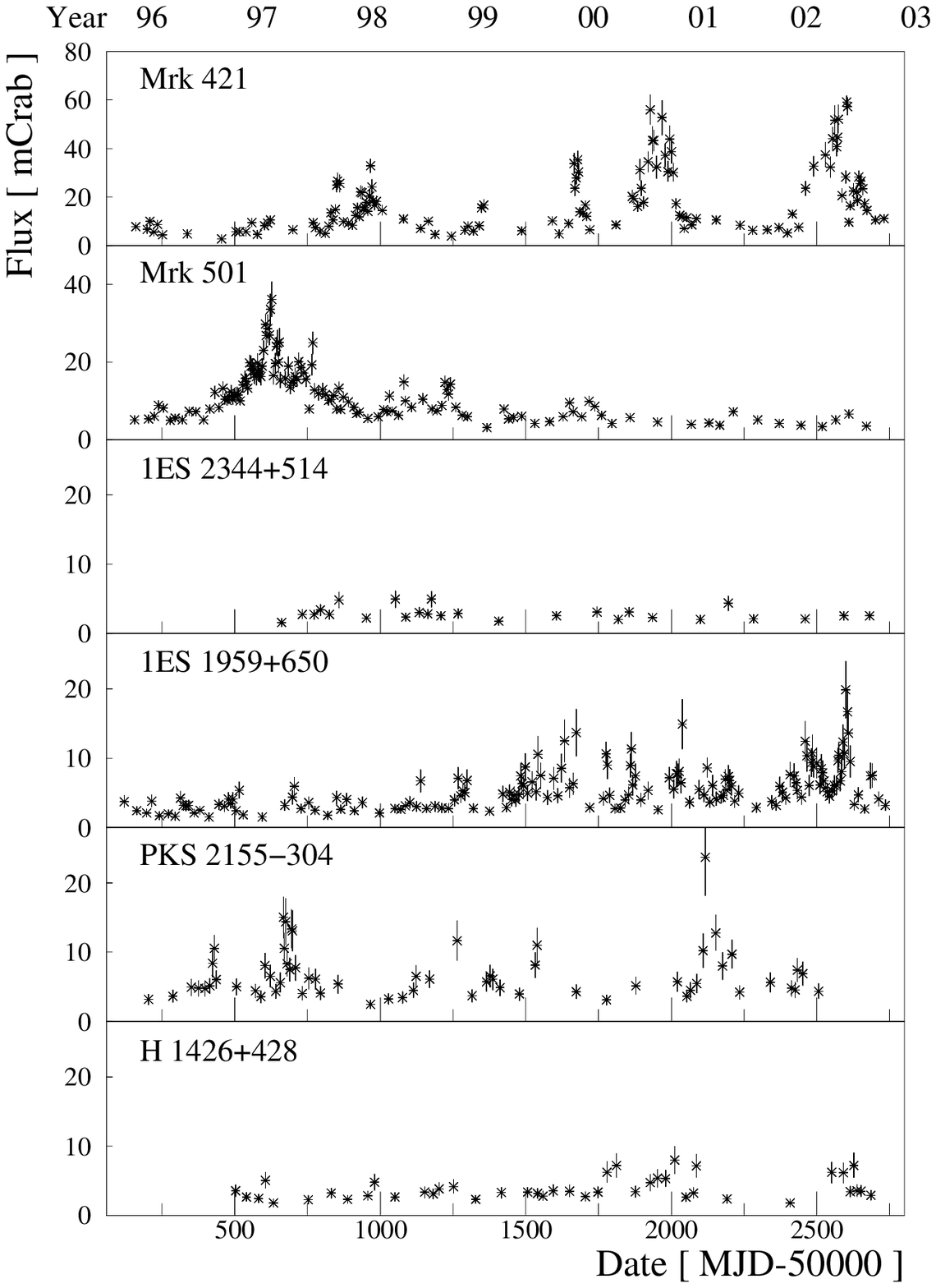}}
\end{center}
%\parbox[b]{55mm}{
\caption{\label{asm} \small
{\it RXTE} All Sky Monitor 2--12 keV lightcurves for the 6 established TeV-blazars. The data have been
binned to assure a certain minimum signal to noise ratio per point (7~$\sigma$ for Mrk 421 and 
Mrk 501, and 4~$\sigma$ for the other sources). For Mrk 421, Mrk 501, 1ES~1959+650, and 
PKS~2155-304 prolonged phases of strong flaring activity can be recognized.}
\end{figure}

In the case of TeV blazars, the large detection area of Cherenkov 
telescopes of several times 10$^5$ m$^2$ 
makes it possible to assess $\gamma$-ray flux variations on time scales of minutes.
As the keV X-ray and TeV $\gamma$-ray emission from these sources is probably
produced by electrons of overlapping energy ranges as synchrotron
and Inverse Compton emission, respectively, observations
of rapid flux and spectral variability in both bands complement 
each other ideally. 
The observations can thus be used to constrain and, in principal, even over-constrain models.
More specifically, the X-ray and TeV $\gamma$-ray observations yield a
measurement of the jet Doppler factor $\delta_{\rm j}$ and the
jet magnetic field $B$ at the jet base.
Observations of TeV blazars can thus reveal key information about the 
astrophysics of mass-accretion onto supermassive black holes 
and the formation of AGN jets.
Unfortunately, the interpretation of the TeV $\gamma$-ray data is not
unambiguous owing to the highly uncertain extent of extragalactic absorption 
of TeV $\gamma$-rays in pair-production processes with photons of the 
Cosmic Infrared Background (CIB) and the Cosmic Optical Background (CIB).
Although X-ray and $\gamma$-ray observations of TeV-blazars might 
ultimately be used to measure the CIB/COB, 
a considerable number of sources is needed as it is difficult 
to disentangle source physics and CIB/COB absorption for individual sources
\cite{Bedn:99,Copp:99,Kraw:02}.\\[2ex]
Owing to its hard X-ray synchrotron emission and low redshift ($z\,=$ 0.047), 
the BL Lac object 1ES~1959+650 had long been considered a prime-candidate
TeV $\gamma$-ray source (e.g.\ Stecker, De Jager \& Salamon 1996, Costamante \& Ghisellini 2002).
The ``Utah Seven Telescope Array'' collaboration reported the detection 
of TeV $\gamma$-ray emission from the source with a total statistical 
significance of 3.9~$\sigma$ \cite{Nish:99}. 
The average flux measured during the 1998 observations 
was about that from the Crab Nebula.
Motivated by the X-ray properties, the Telescope Array detection, and a 
tentative detection of the source by the HEGRA Cherenkov telescopes in 2000 and 2001,
we proposed pre-approved pointed {\it RXTE} target of opportunity observations. 
These observations were to take place immediately after a predefined increase in the
X-ray or gamma-ray activity was detected with the {\it RXTE} All Sky Monitor (ASM) 
or the Whipple 10 m Cherenkov telescope.
Following the detection of a spectacular TeV $\gamma$-ray flare on May 17, 2002
with the Whipple 10~m telescope by the VERITAS  {\it (Very Energetic Radiation Imaging Telescope 
Array System)} collaboration we invoked the pointed {\it RXTE} 
observations as well as simultaneous observations
in the radio, optical, and TeV $\gamma$-ray bands.
The Whipple  \cite{Hold:03} and HEGRA \cite{Ahar:03b} data showed 
that the $\gamma$-ray flux was strongest during the 
first 20 days of observations with peak fluxes of between 4 and 5 Crab units; 
subsequently, the flare amplitude decreased slowly.
Following Mrk 421 ($z\,=$ 0.031) and Mrk 501 ($z\,=$ 0.034), 1ES~1959+650
is now the third TeV $\gamma$-ray blazar with a high-state flux much stronger than 
that from the Crab Nebula, allowing us to measure the $\gamma$-ray lightcurve on a time 
scale of a couple of minutes, and to take energy spectra with good photon statistics
on a nightly basis.
Since the discovery of the first TeV blazar Mrk 421 in 1992 \cite{Punc:92}, the number of
well established blazars has now grown to 6 (see Table \ref{blazars}). 
Fig.\ \ref{asm} shows the 2-12 keV flux from these 6 sources as measured in the years
1996 to 2003 with the {\it RXTE} ASM. For Mrk 421, Mrk 501, 1ES~1959+650, and PKS~2155-304 long 
flaring phases extending over several weeks can be recognized. 
While Mrk 421, 1ES~1959+650, and PKS~2155-304 flare frequently, 
Mrk 501 flared in 1997, but showed only modest fluxes thereafter.
The prolonged flaring phases offer ideal opportunities to study these objects
with high photon statistics.\\[2ex]
In this paper, we discuss the results of the 2002 multiwavelength campaign on 1ES~1959+650.
We present new radio, optical and {\it RXTE} X-ray data taken between May 16, 2002 and August 14, 
2002, and combine these data with the already published Whipple and HEGRA TeV $\gamma$-ray data.
In Sect.\ \ref{data} we present the data sets and the data reduction methods. 
In Sect.\ \ref{overview} we give an overview of the combined light curves, and in  
Sect.\ \ref{lightcurves} we scrutinize certain episodes of the light curves in more detail. 
After discussing the flux correlations in different energy bands in Sect.\ \ref{sed},
we present the radio to $\gamma$-ray SEDs of 1ES 1959+650 and show results of initial 
modeling with the data in Sect.\ \ref{seds}.
With the addition of another TeV blazar with good broadband data,
we consider the set of all TeV blazars to begin to look for a connection
of the jet properties to the properties of the central engine in Sect.\ \ref{bhm}.
We discuss the implications of our observations in Sect.\ \ref{disc}.

We use the following cosmological parameters $H_0\,=$ $h_0\times$ 100~km~s$^{-1}$~Mpc$^{-1}$ 
with $h_0\,=0.65$, $\Omega_{\rm M}\,=$ 0.3, and $\Omega_{\rm \Lambda}\,=$ 0.7.
The redshift of 1ES~1959+650 translates into a luminosity distance of 229.5~Mpc. 
Errors on the best-fit results of $\chi^2$-fits to the {\it RXTE} data are
given on the 90\% confidence level.
All other errors are quoted on the 1~$\sigma$ confidence level.
\section{Data Sets and Data Reduction}
\label{data}
\subsection{Radio Observations}
We used the University of Michigan 26-meter paraboloid to monitor 1ES~1959+650 
at 4.8 GHz and 14.5 GHz between May 5 and August 9, 2002. 
Each observation consisted of a series of ON-OFF measurements 
taken over a 30-40 minute time period. All observations were made within a total hour 
angle range of about 5 hours centered on the meridian. 
The calibration and reduction procedures have been described in \cite{Alle:85}.
Some daily observations were averaged to improve the signal-to-noise 
ratio.

Additional flux density measurements were made with the Very Large Array (VLA) 
of the National Radio Astronomy Observatory 
\footnote{The NRAO is a facility of the  National Science Foundation operated 
under cooperative agreement by Associated Universities, Inc.} 
at frequencies of 43.315 GHz, 22.485 GHz, 14.965 GHz, 8.435 GHz, and 4.885 GHz
on May 7 and June 7, 2002, in snap-shot mode (single scans). 
Observations were made in A-array configuration on May 7 and B-Array configuration June 7. 
Pointing checks were incorporated in an effort to keep 
the program sources near the centers of the primary antenna beams. 
The data were reduced within the AIPS software package supplied by NRAO, 
following the standard procedures outlined in the AIPS Cookbook. 
Flux density calibration was accomplished through observations of the source
3C~286. Some of the flux density measurements at the highest three frequencies 
were discarded owing to erratic variations among antenna pairs. 
Several secondary calibration sources were used for checks on the 
final flux density scale.
\subsection{Optical Observations}
We present two optical data sets.
One was taken with the 0.4~m telescope at Boltwood Observatory
(Stittsville, Ontario) between May 18 and August 14, 2002, using V, R, and I
broadband filters. 
The aperture photometry was performed with custom software and used the 
comparison star 4 from Villata et al. (1998). 
Data points were obtained from averaging over between 4 and 6 
2-minute exposures. 
Relative V and R band magnitudes were converted to absolute magnitudes using
the published absolute magnitudes from Villata et al. (1998).
We are not aware of a published measurement of the absolute I magnitude of star 4 
and we give the results only as relative magnitudes 
mag(1ES~1959+650)-mag(star~4).
The typical statistical error on the relative photometry of each data point is 
0.02 mag. The absolute photometry has an additional error of 0.03 mag.

The other data set was taken with the 0.7 m telescope at 
the Abastumani Observatory in Georgia from May 19 to July 12, 2002, using
an R filter for all observations. The frames were reduced using \verb+DAOPHOT II+.
The absolute magnitude of 1ES~1959+650 was determined by comparison with the 
standard stars 4, 6, and 7 from Villata et al. (1998).
In 20 nights, 192 measurements of 5 minutes exposure time were taken.
The statistical error on the relative photometry is 0.1 mag.
The absolute photometry has an additional error of 0.05 mag.
For both data sets we did not attempt to remove the light contribution from the
host galaxy.
\subsection{X-Ray Observations}
The X-ray analysis was based on the 3-25~keV data from the Proportional Counter Array 
(PCA; Jahoda et al. 1996)  on board the {\it RXTE} satellite. 
Standard-2 mode PCA data gathered with the top
layer of the operational Proportional Counter Units (PCUs) were
analyzed.  The number of PCUs operational during a pointing varied between
2 and 4. We did not use the 15--250 keV data from the High-Energy X-ray Timing 
Experiment HEXTE (Rothschild et al. 1998) owing to their poor signal to noise ratio. 

After applying the standard screening criteria and removing by hand
abnormal data spikes, the net exposure in each
Good Time Interval ranged from 160 secs to 4.43 ksecs (see
Table~\ref{xflux}). Spectra and lightcurves were extracted with
\verb+FTOOLS+~v5.1\verb+A+. Background models were generated with the tool
\verb+pcabackest+, based on the {\it RXTE} GOF calibration files for a
``bright'' source  with more than 40 counts/sec. 
Comparison of the background models and the data at energies above 30 keV
showed that the model underestimated the background by 10\%.
We corrected for this shortcoming by scaling the background model with a correction factor
of 1.1. Response matrices for the PCA data were created with the 
script \verb+pcarsp+~v.7.11.

The spectral analysis was performed with the \verb+Sherpa+~v.2.2.1
package. A galactic neutral hydrogen column density of 1.027~$\rm
\times~10^{21}~cm^{-2}$ was used for all observations. 
Since the analysis is restricted to the energy region above 3~keV the 
hydrogen column density has only
a very minor influence on the estimated model parameters. 
Single power-law models resulted in statistically acceptable fits for all
data sets.
\subsection{Gamma-Ray Observations}
\label{gamma}
1ES~1959+650 was monitored on a regular basis as part of the BL Lac program at the
Whipple Cerenkov telescope during the 2001/2002 observing season; it
was during these observations that 1ES~1959+650 was seen to go into
an active state. 
Following the detection of strong flares on May 16 and 17, 2002, we 
coordinated simultaneous observations of 1ES 1959+650 with
the Whipple and HEGRA Cherenkov telescopes. 
The observations with the Whipple 10~m Cherenkov Telescope
began on May 16, 2002 and ended on July 8, 2002 \cite{Hold:03}.
The total data set consists of 39.3 hrs of ON-source data, together 
with 7.6 hrs of OFF-source data for background comparison. 
The Whipple telescope is located in southern Arizona, USA, on Mt.\ Hopkins and is part
of the Whipple observatory.
At this latitude,  1ES 1959+650 culminates at a zenith angle of 33.5$^\circ$, and so the data 
were necessarily taken at large zenith angles, between 33.5$^\circ$  and  53.5$^\circ$. 
The data were corrected for large zenith angles and for a temporary reduction of 
the telescope detection efficiency using the method of LeBohec and Holder (2003) 
which involves measuring the response of the telescope to cosmic rays.
While correcting the $\gamma$-ray detection rates for the reduced 
telescope sensitivity is straightforward, energy spectra can not be determined 
with the standard tools and further studies of the Whipple energy spectra
are underway. The peak energy\footnote{The peak energy is defined as the energy at which the 
differential $\gamma$-ray detection rate peaks, assuming a source with the same 
$\gamma$-ray spectrum as the Crab Nebula.} lies at about 600 GeV for the majority of observations.

Motivated by the HEGRA detection of the source in 2000 and 2001, as well as by the
strong flaring activity in May, 2002, the HEGRA system of five Cherenkov telescopes
(Canary Island La Palma) regularly monitored 
1ES~1959+650 in 2002. A total of 89.6 hrs of data were taken
during moonless nights from May 18 to September 11, 2002 (Aharonian et al.\ 2003b).
Typically, each night comprises about 1 hour of observation time around the
object's culmination. Owing to the declination of 
1ES~1959+650, the object could only be observed at zenith angles 
above 35.9$^\circ$ leading to a mean peak energy of 1.4 TeV. 
All observations were carried out in the so-called wobble mode allowing 
for a simultaneous measurement of the background rate induced by 
charged cosmic rays. 
The HEGRA collaboration determined the differential 1.3 TeV - 12.6 TeV 
energy spectrum 
$dN/dE\,=$ $N_0$ $(E/\rm 1\,TeV)^{-\Gamma}$ of 1ES~1959+650 for a high-flux 
data set and a low-flux data set. 
The high-flux data set used all 2002 data for which the diurnal integral flux above 2 TeV
surpassed that from the Crab and gave
$N_0\,=$ (7.4$\pm$1.3$_{\rm stat}$$\pm$0.9$_{\rm syst}$) 10$^{-11}$ photons cm$^{-2}$ s$^{-1}$
TeV$^{-1}$, and $\Gamma\,=$ (2.83$\pm$0.14$_{\rm stat}$$\pm$0.08$_{\rm syst}$).
The low-flux data set used all 2000-2002 data for which the diurnal integral flux above 2 TeV
was less than 0.5 Crab units and gave
$N_0\,=$ (7.8$\pm$1.5$_{\rm stat}$$\pm$1.0$_{\rm syst}$) 10$^{-12}$ photons cm$^{-2}$ s$^{-1}$
TeV$^{-1}$, and $\Gamma\,=$ (3.18$\pm$0.17$_{\rm stat}$$\pm$0.08$_{\rm syst}$).

In the following, we quote integral $\gamma$-ray flux in Crab units above energy thresholds
of 600~GeV and 2~TeV, for the Whipple and HEGRA data points, respectively.
In the case of HEGRA, the analysis threshold has been chosen well above the peak energy to
minimize systematic uncertainties in the region of the trigger threshold.
The normalization of the fluxes in Crab units renders the results
largely independent of Monte Carlo simulations. The drawback of the method is that 
different energy thresholds can introduce flux offsets if the source energy spectrum deviates from the
Crab energy spectrum. Based on the HEGRA results on the correlation
of the $\gamma$-ray flux level and $\gamma$-ray photon index, 
we estimate that these offsets are smaller than 20\% for $>600$~GeV 
flux levels on the order of 1 Crab and higher, and smaller than a 
factor of 2 for flux levels well below 1 Crab. 
Based on the Whipple measurement of the energy spectrum from the Crab nebula
\cite{Hill:98}, a flux of 1 Crab corresponds to a differential 1 TeV flux
of (3.20$\pm$0.17$_{\rm stat}$$\pm$$0.6_{\rm syst}$) $\times10^{-11}$ photons
cm$^{-2}$ s$^{-1}$ TeV$^{-1}$ and to a $\nu F_\nu$ flux of 
(5.12$\pm$0.27$_{\rm stat}$$\pm$$0.96_{\rm syst}$) $\times10^{-11}$ 
ergs cm$^{-2}$ s$^{-1}$.
\section{Results of the Multiwavelength Campaign}
\label{overview}
\begin{figure}[bh]
\begin{center}
\resizebox{12cm}{!}{
\plotone{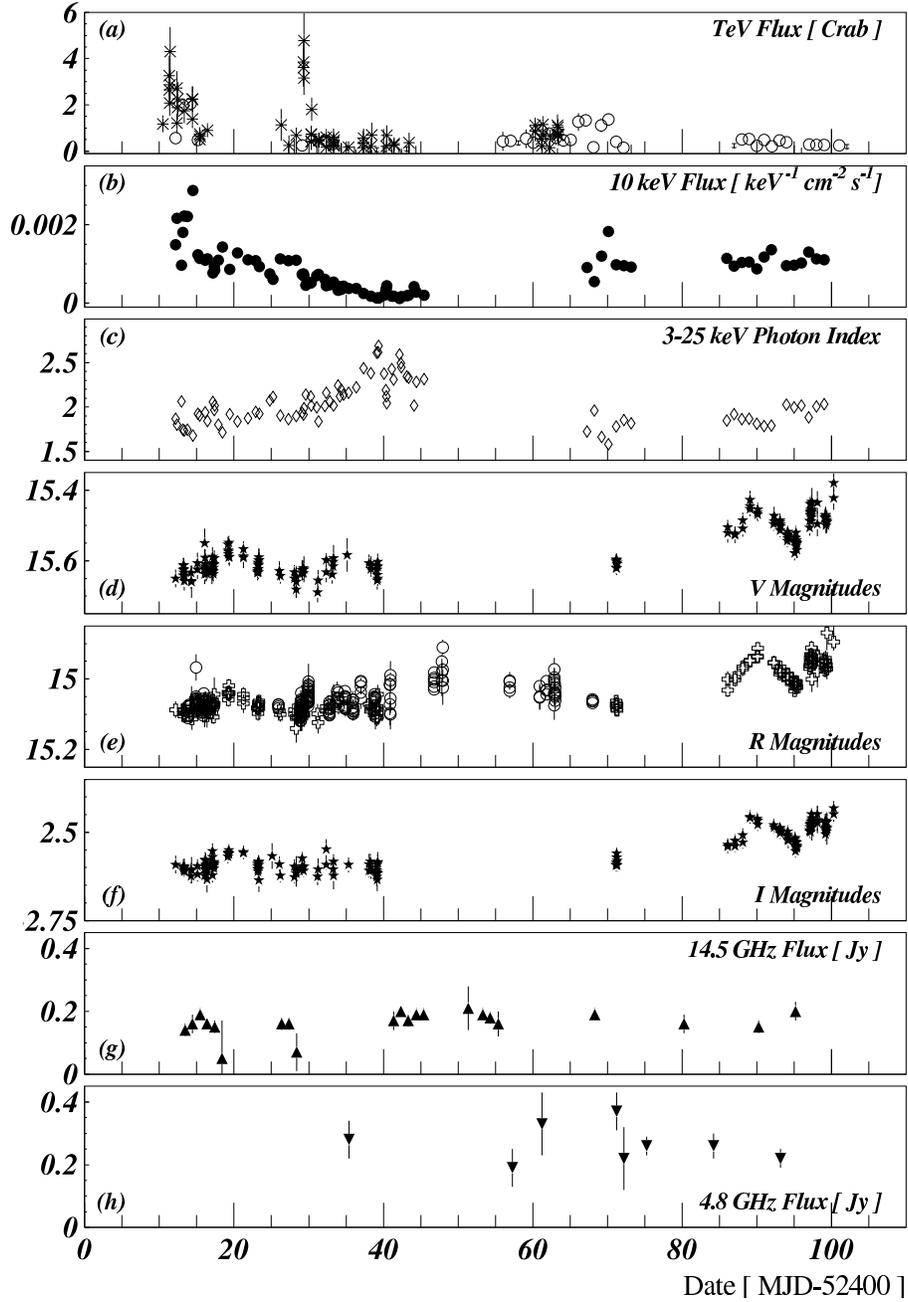}}
\end{center}
%\parbox[b]{55mm}{
\caption{\label{all} \small
Results from the 1ES~1959+650 multiwavelength campaign (May 16, 2002 -- August 14, 2002):
(a) Whipple (stars) and HEGRA (circles) 
integral TeV $\gamma$-ray fluxes in Crab units above 600 GeV and 2 TeV, respectively;
the Whipple data are binned in 20 min bins, and the HEGRA data in diurnal bins;
(b)  {\it RXTE} X-ray flux at 10 keV;
(c)  {\it RXTE} 3-25 keV X-ray photon index;
(d)  absolute V magnitudes (Boltwood);
(e)  absolute R magnitudes (swiss crosses: Boltwood, circles: Abastumani);
(f)  relative I magnitudes (Boltwood);
(g) the 14.5 GHz flux density (UMRAO) and 
(h) the 4.8 GHz flux density (UMRAO).
}
\end{figure}

%
% TeV 
%
Figure \ref{all} shows from top to bottom the integral TeV flux,
the X-ray flux at 10 keV, the 3-20 keV X-ray photon index, 
the V, R, and I band optical data, and the 14.5 GHz and 4.8 GHz radio data.
The TeV $\gamma$-ray data (Fig.\ \ref{all}a) show several strong flares during the first
20 days of the campaign with a flux surpassing 2 Crab units on
May 17--20 (MJD 52411--52414) and again roughly two weeks later on June 4 (MJD 52429).
Subsequently, the flux leveled off to about 0.3 Crab units with the exception of
two flares on July 11--12 (MJD 52466-52467) and 
July 14--15 (MJD 52469-52470) with a flux between 1 and 1.5 Crab units.
Holder et al.\ (2002) studied the fastest $\gamma$-ray flux variability time scales 
based on the Whipple data and found a rapid flux increase with an $e$-folding time of 10 hrs.
The large ``gaps'' in the $\gamma$-ray lightcurves originate from the fact 
that the Cherenkov telescopes are operated during moonless nights only.

The 10 keV X-ray  flux (Fig.\ \ref{all}b, Table \ref{xflux}) 
was strongest on May 18-20 (MJD 52412-52414). 
It slowly decreased by a  factor of 18.7 from the maximum on May 20 
to a minimum on June 17 (MJD 52442).
As we will discuss in more detail in the next section, the TeV $\gamma$-ray
and X-ray fluxes seem to be correlated, with the notable exception of an ``orphan''
TeV $\gamma$-ray flare on June 4, 2002 (MJD 52429) that is not associated with 
increased X-ray activity.
From July 17 (MJD 52469) until the end of the campaign, 
the X-ray flux stayed at a consistently high level: a factor of 1.7
below the maximum flux observed at the beginning of the campaign, 
and a factor of 11.5 above the minimum flux measured on June 17 (MJD 52442).
The TeV emission level during this ``X-ray plateau state'' is about a factor of
2 lower than at similar X-ray flux levels earlier in the campaign.

We analyzed the X-ray flux variability time scale by computing the $e$-folding times
from the flux changes between observations:
$\tau\,=$ $\Delta\,t\,/$ $\Delta\,ln\, F(10\rm \, keV)$   
with $\Delta\,t$ being the time difference between two observations and 
$\Delta\, ln\, F(10\rm \, keV)$  is the difference of the 
logarithms of the 10~keV fluxes.
The shortest $e$-folding times are given in Table~\ref{inc}.
We detected faster flux increases than flux decreases: the fastest flux increase
has an $e$-folding time of $\simeq$5.9~hrs; the fastest flux decrease
has an $e$-folding time of $\simeq$15.2~hrs.

The 3-25 keV photon index $\Gamma$ ($dN/dE\propto E^{-\Gamma}$)
(Fig.\ \ref{all}c, Table \ref{xflux}) varies between 1.6 and 2.4. 
The X-ray photon index and the X-ray flux are clearly correlated, 
higher flux corresponding to harder energy spectrum.
Values well below and well above the value of $\Gamma\,=$ 2 
show that the low-energy (presumably synchrotron) component 
sometimes peaked above 10 keV, and sometimes below 10~keV.
We searched for rapid spectral changes by analyzing photon index variations
between {\it RXTE} observations, see Table~\ref{hard}.
The photon index $\Gamma$ decreased (spectral hardening) by up to 0.09 hr$^{-1}$ 
and increased (spectral softening) by up to 0.04 hr$^{-1}$.
As a consequence of synchrotron cooling which is more efficient at higher energies,
leptonic models predict that the X-ray emission is harder during the rising 
phase of a flare than during its decaying phase \cite{Kard:62}.
Careful inspection of the X-ray lightcurve and the photon indices does not show 
evidence for such a behavior.
While the detection of this effect would impose a constraint on the jet magnetic field
and the Doppler factor, the non-detection allows large 
regions in the $\delta_{\rm j}$ - $B$ plane \cite{Kraw:02}.

The V,R, and I band optical data (Fig.\ \ref{all}d-f) show
flux variations of about 0.1 mag on typical time scales of about 10 days.
Remarkably, the mean optical brightness increased from the first 4 weeks to the last 2
weeks of the campaign by about 0.1 mag in all three optical bands. 
Both, the optical and the X-ray fluxes increased slowly during the campaign. 
Apart from this joint slow flux increase, we did not find any evidence for a 
correlation between the optical and the X-ray or the TeV $\gamma$-ray fluxes.
We searched for optical intra-day flux variability by fitting models to the 
data of individual days.
Although we performed very long observations of up to $\simeq$7 hrs per night with
small statistical errors of 0.02 mag per 10 minute exposure time, the reduced chi-square values 
did not show any evidence for statistically significant
intra-day flux variability.

Based on the diurnal brightness averages in the three bands,
we computed fastest rise and decay $e$-folding times 
of 0.07 mag day$^{-1}$ and 0.03 mag day$^{-1}$,
respectively. Within the statistical errors, the V-R and V-I colors stay constant 
throughout the full campaign.

The 14.5 GHz and 4.8 GHz radio data taken with the UMRAO (Fig.\ \ref{all}g-h) 
do not show significant flux variations. 
A fit of a constant flux level to the 14.5 GHz data gives 
a mean of 0.174$\pm$0.004~Jy with a chi-square value of 
24.5 for 21 degrees of freedom (chance probability of 27\%).
The mean flux is consistent with the 14.965 GHz flux of 
0.18$\pm$0.01~Jy measured with the VLA on May 7.
A fit of a constant flux level to the 4.8 GHz UMRAO data gives 
a mean of 0.254$\pm$0.016~Jy with a chi-square value of 
7.1 for 7 degrees of freedom (chance probability of 42\%).
The mean 4.8 GHz flux is compatible with the 4.885 GHz flux measured on
June 7 with the VLA of 0.23$\pm$0.01~Jy, and with 4.85 GHz values of 
0.253$\pm$0.023~Jy and 0.246$\pm$0.037~Jy reported by
Gregory \& Condon (1991) and Becker \& White (1991), respectively.
\section{Detailed Lightcurves}
\label{lightcurves}
In this section we discuss the lightcurves in more detail 
by dividing the data into 4 epochs (Epoch 1: MJD 52410--52419, 
Epoch 2: MJD 52420--52445, Epoch 3: MJD 52460--52474, Epoch 4: MJD 52486-52500).
In the following figures, we show only 
the observational bands where a substantial number of data points 
were recorded.

\begin{figure}[bh]
\begin{center}
\resizebox{12cm}{!}{
\plotone{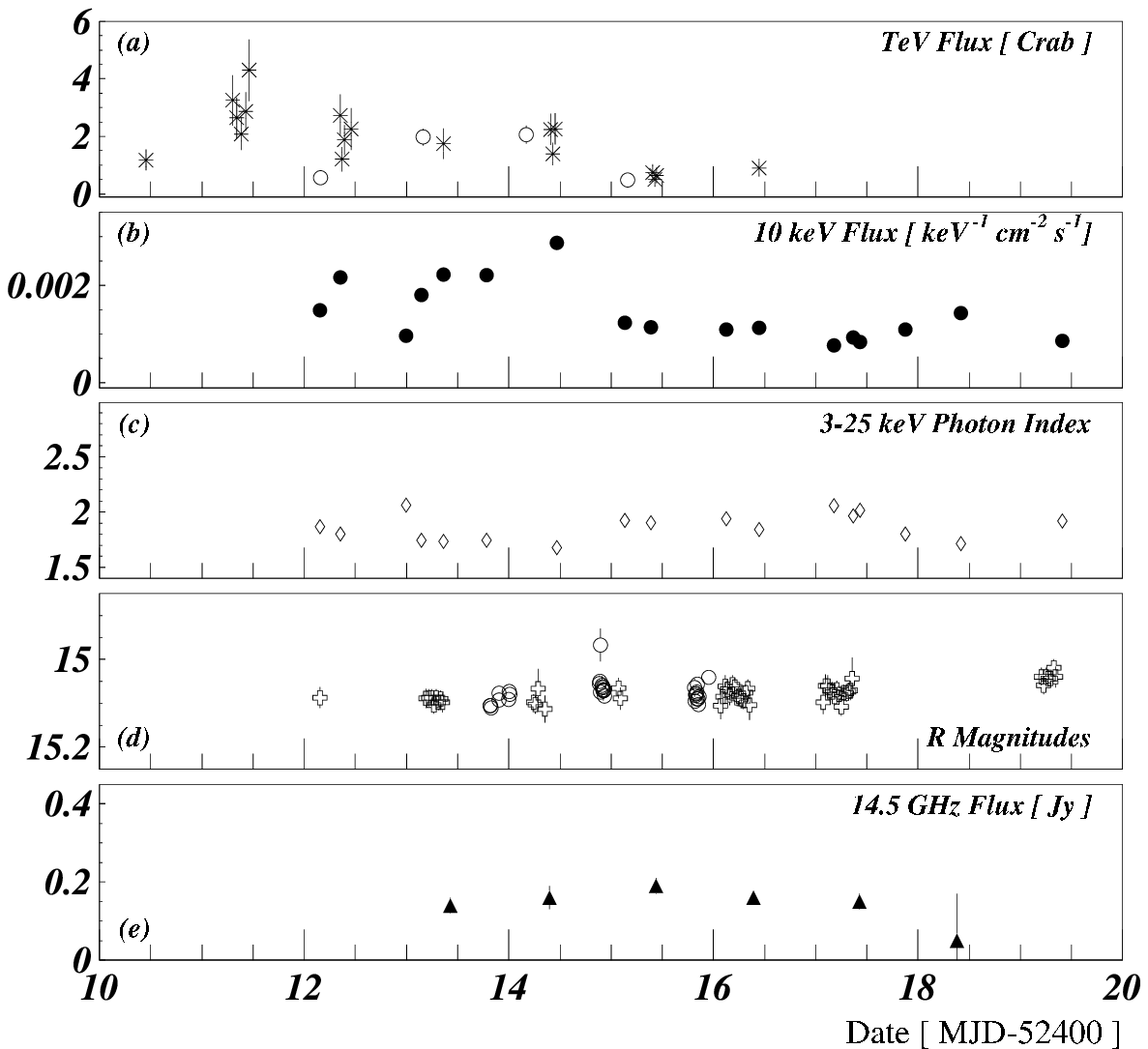}}
\end{center}
%\parbox[b]{55mm}{
\caption{\label{epoch1} 1ES~1959+650 data from Epoch 1 of the campaign (symbols as in Fig.\ \ref{all}).}
\end{figure}

Fig.\ \ref{epoch1} shows the data from Epoch~1 (May 16--25, MJD 52410--52419). 
Note that the X-ray observations started on May 18, UTC 3:26, within less
than 24 hrs of the initial detection of strong $\gamma$-ray flaring activity 
from 1ES~1959+650 with the Whipple 10~m telescope.
The $\gamma$-ray and X-ray fluxes seem to be correlated, both showing a strong flux 
increase on May 18 (MJD 52412) and a strong flux decrease on May 21 (MJD 52415).
From May 19 to May 20, the X-ray flux increases by 20\% without a similar increase
in the $\gamma$-ray band.
While the source brightened in the time interval May 19--25 (MJD 52413.27--52419.40) 
by about 0.1 mag in all three optical bands, the X-ray flux decreased by a factor 
of $\sim$3 over the same time interval.

\begin{figure}[bh]
\begin{center}
\resizebox{12.2cm}{!}{
\plotone{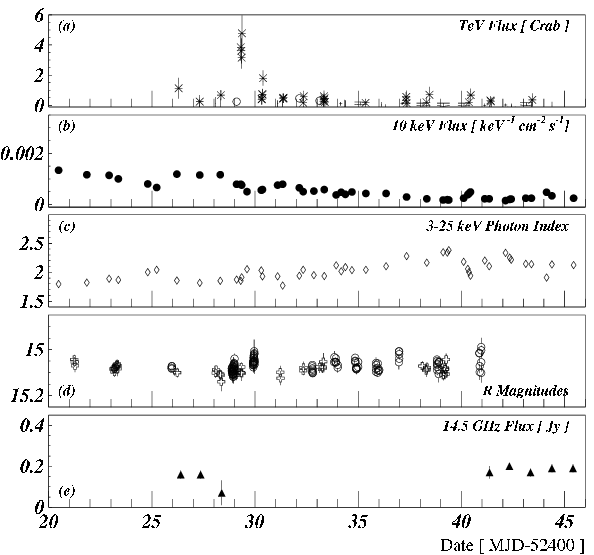}}
\end{center}
%\parbox[b]{55mm}{
\caption{\label{epoch2} 1ES~1959+650 data from Epoch 2 of the campaign (symbols as in Fig.\ \ref{all}).}
\end{figure}

\begin{figure}[bh]
\begin{center}
\resizebox{12cm}{!}{
\plotone{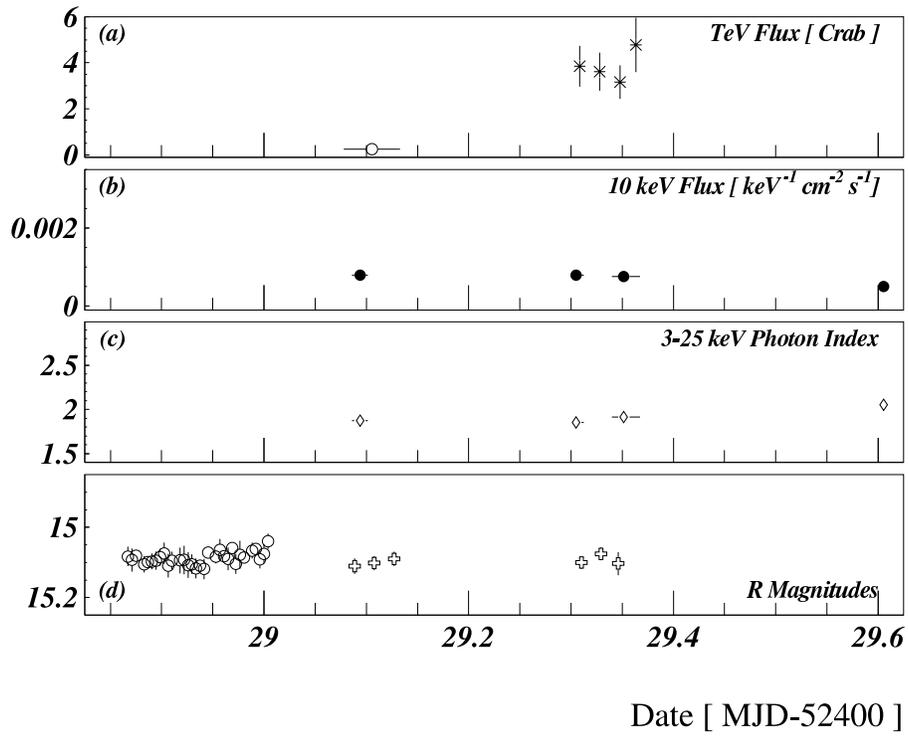}}
\end{center}
%\parbox[b]{55mm}{
\caption{\label{orphan} 1ES~1959+650 data showing the ``orphan'' $\gamma$-ray flare observed on June 4, 2002
(symbols as in Fig.\ \ref{all}).}
\end{figure}

The data from Epoch 2 (May 26--June 21, MJD 52420--52446)
are shown in Fig.\ \ref{epoch2}.
Except for one strong $\gamma$-ray flare, the $\gamma$-ray flux stayed well below 2 Crab units.
The X-ray flux decayed slowly, and the optical brightnesses in the three bands
meandered around their mean values by 0.05 mag. 
The most interesting feature of the full observation campaign is the strong ``orphan'' 
$\gamma$-ray flare on June 4 (MJD 52429.308--52429.362) which Fig.\ \ref{orphan} shows in more detail.
While HEGRA measured a low flux of 0.26$\pm$0.21 Crab units on MJD 52429.106,
the Whipple observation 5 hrs later revealed a high flux of 4 Crab units.
The X-ray flux (measured at the same time as the $\gamma$-ray data) 
did not show any sign of an increased activity: the 10 keV flux stayed constant
and later even decreased compared to the observation taken 5~hrs earlier.
Similarly, the X-ray photon index and the optical magnitudes do not show
any irregularity during the $\gamma$-ray flare.

\begin{figure}[bh]
\begin{center}
\resizebox{12cm}{!}{
\plotone{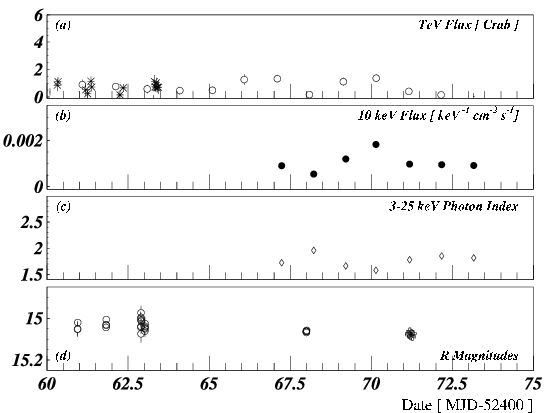}}
\end{center}
%\parbox[b]{55mm}{
\caption{\label{epoch3} 1ES~1959+650 data from Epoch 3 of the campaign (symbols as in Fig.\ \ref{all}).}
\end{figure}

The results from Epoch 3 (July 5--19, MJD 52460--52474) are presented in Fig.\ \ref{epoch3}.
The $\gamma$-ray and X-ray fluxes show a very similar development 
with joint flux minimums on July 13 (MJD 52468) and July 18 (MJD 52473), 
and a joint flux maximum on July 15 (MJD 52470). As during the full campaign, 
the X-ray flux and X-ray photon index are tightly correlated. 

\begin{figure}[bh]
\begin{center}
\resizebox{12cm}{!}{
\plotone{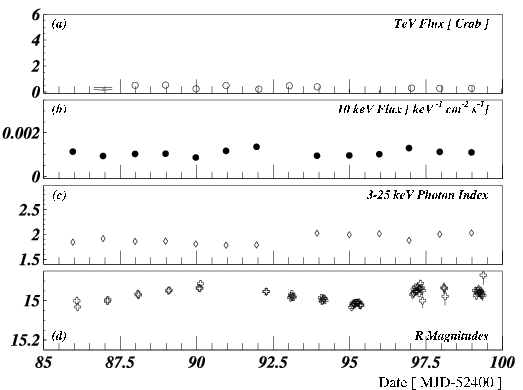}}
\end{center}
%\parbox[b]{55mm}{
\caption{\label{epoch4} 1ES~1959+650 data from Epoch 4 of the campaign (symbols as in Fig.\ \ref{all}).}
\end{figure}

The data from Epoch 4 (July 31--August 14, MJD 52486--52500) are shown in Fig.\ \ref{epoch4}.
While statistical errors hamper the interpretation of the $\gamma$-ray data,
the X-ray flux varied by 50\% and the optical flux by 0.1 mag.
\section{Flux Correlations in Different Energy Bands and X-Ray 
Hardness--Intensity Correlation}
\label{sed}
\begin{figure}[bh]
\begin{center}
\resizebox{8cm}{!}{
\plotone{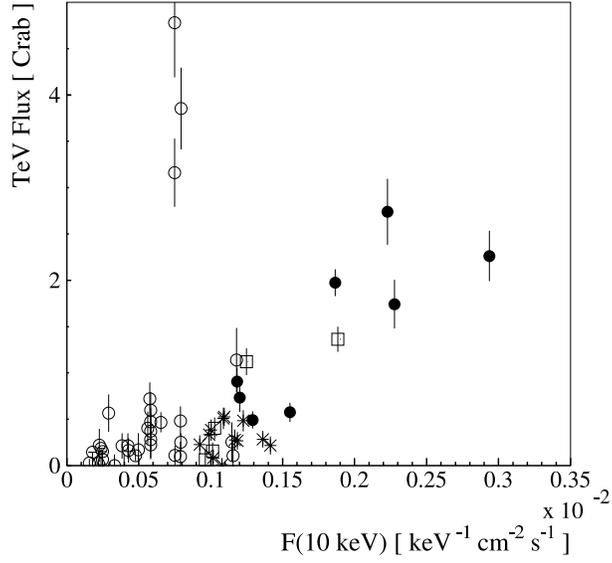}}
\end{center}
%\parbox[b]{55mm}{
\caption{\label{xgcorr} Correlation between the X-ray flux and the Whipple and HEGRA 
$\gamma$-ray fluxes. Epoch~1: full circles, Epoch~2: open circles, Epoch~3: open squares, 
and Epoch~4: asterisks. Only points with a direct overlap of
the $\gamma$-ray and X-ray observations have been included in this graph.}
\end{figure}

\begin{figure}[bh]
\begin{center}
\resizebox{8cm}{!}{
\plotone{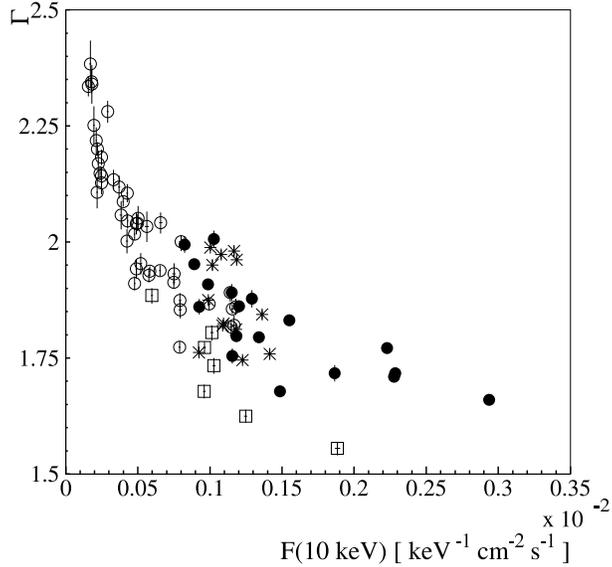}}
\end{center}
%\parbox[b]{55mm}{
\caption{\label{hardness} Correlation between the X-ray flux and the 3-25 keV photon index.
Epoch~1: full circles, Epoch~2: open circles, Epoch~3: open squares, 
and Epoch~4: asterisks.}
\end{figure}

The correlation between simultaneously measured $\gamma$-ray and X-ray fluxes during 
the full campaign
is shown in Fig.\ \ref{xgcorr}; even though the fluxes seem to be correlated in general,
the orphan flare clearly deteriorates the quality of the correlation.
During the observation campaign, the 3-25 keV X-ray photon index and the 10 keV
flux were tightly correlated (Fig.\ \ref{hardness}).
Higher flux levels are accompanied by harder energy spectra, as is typical
for BL Lac objects.
The photon-index/flux correlation shows some slow evolution during the 
multiwavelength campaign with some exceptionally hard energy spectra 
recorded during Epoch~3, around July 15 (MJD 52470).
In the hardness--intensity plane we did not detect clockwise or anti-clockwise 
loops during flares. Such loops are expected to occur as a consequence of diffusive 
particle acceleration at strong shocks and synchrotron cooling of the radiating particles 
(electrons or protons) \cite{Taka:96,Kirk:99}.
The sparse observational sampling might be responsible for our non-detection.
\section{Spectral Energy Distribution and SSC Modeling}
\label{seds}
The X-ray and $\gamma$-ray emission from TeV-Blazars are commonly attributed to the SSC 
mechanism in which a population of high-energy electrons emits synchrotron 
radiation, followed by Inverse Compton scattering of synchrotron photons
to TeV energies.

\begin{figure}[bh]
\begin{center}
\resizebox{10cm}{!}{
\plotone{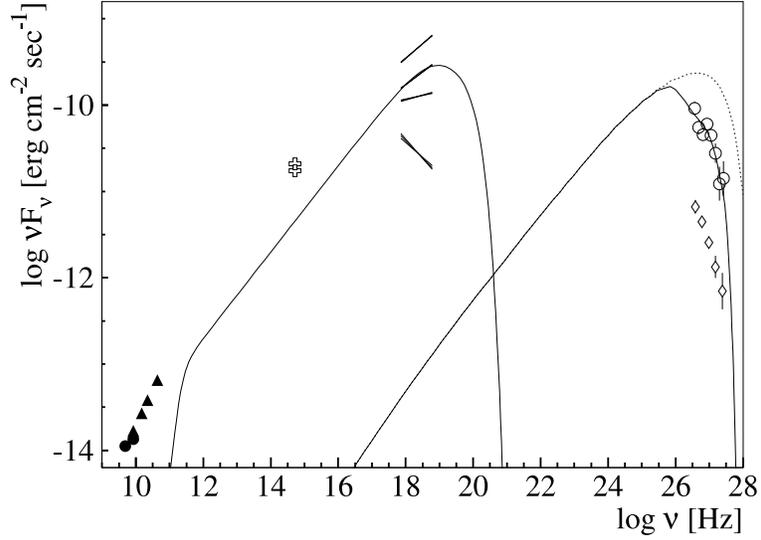}}
\end{center}
%\parbox[b]{55mm}{
\caption{\label{sed01} \small Radio to $\gamma$-ray SEDs of the blazar 1ES 1959+650.
The VLA data were taken on May 7 (filled upward triangles) and June 7, 2002 (filled circles).
The optical R-band data (swiss crosses) shows the minimum and maximum fluxes detected 
during the full multiwavelength campaign. 
Four {\it RXTE} energy spectra are given (results from the single power-law fits); 
from top to bottom we show: 
(i) the spectrum during a strong and spectrally hard flare observed on May 20, 
(ii) an estimate of the time averaged spectrum corresponding to the HEGRA ``high-state'' 
energy spectrum; (iii) the spectrum measured during the ``orphan'' $\gamma$-ray flare
on June 4; (iv) the spectrum of the {\it RXTE} pointing with the steepest photon index (June 14).
The open circles show the HEGRA ``high-state'' energy spectrum measured during
6 nights with a $>$2 TeV integral flux above 1 Crab unit, and the diamonds show the HEGRA 
``low-state'' energy spectrum acquired during all 2000-2002 nights with an integral  flux 
of less than 0.5 Crab units. 
An SSC model of the high-state HEGRA data and the corresponding high-state {\it RXTE} data
is shown by the solid line; the dotted line shows the model before correction for
extragalactic absorption. 
The model parameters are: $\delta_{\rm j}\,=$ 20, $B\,=$ 0.04~G, 
$R\,=$ 5.8$\times$10$^{15}$~cm, log$(E_{\rm min}/\rm eV)\,=$ 3.5, 
log$(E_{\rm max}/\rm eV)\,=$ 12.3,  log$(E_{\rm b}/\rm eV)\,=$ 11.8,
$p_1\,=$ 2, $p_2\,=$ 3, electron energy density 0.22 erg cm$^{-3}$.
}
\end{figure}

In Fig.\ \ref{sed01} we show the radio to $\gamma$-ray SED of 1ES 1959+650 together with
a simple one-zone SSC model.
The model (Krawczynski 2003b, see Inoue \& Takahara (1996), 
and Kataoka et al.\ (1999) for similar codes) assumes a spherical emission 
volume of radius $R$, that moves with bulk Lorentz factor $\Gamma$ toward the observer.
The radiation is Doppler shifted by the Doppler factor
\begin{equation}
\delta_{\rm j}\,=\,\left[\Gamma(1-\beta\,\cos{(\theta)})\right]^{-1},
\end{equation}
with $\beta$ the bulk velocity of the plasma in units of the speed of light, 
and $\theta$ the angle between jet axis and the line of sight in the observer frame.
The emission volume is filled with an isotropic electron population and a 
randomly oriented magnetic field $B$.
We assume that the energy spectrum of the electrons in the jet frame can be described 
by a broken power law with low-energy ($E_{\rm min}$ to $E_{\rm b}$)
and high-energy ($E_{\rm b}$ to $E_{\rm max}$) indices
$p_1\,=$ 2 and $p_2\,=$ 3, respectively ($p_i$ from $dN/d\gamma\,\propto$ $\gamma^{-p_i}$, 
$E$ is the electron energy in the jet frame).
Motivated by the similar SEDs and flux variability time scales of Mrk 501 in 1997 and 
1ES 1959+650 in 2002, we chose parameter values
similar to those inferred for Mrk 501 from time dependent modeling 
of 1997 X-ray and $\gamma$-ray data \cite{Kraw:02}. 
The dotted line shows the model prediction before taking into account extragalactic extinction and
the solid line shows the SED modified by intergalactic extinction as predicted 
by a CIB/COB model with a reasonable shape. We choose the CIB/COB model of Kneiske et al.\ (2002), 
see e.g.\ Primack et al. (2001) and de Jager \& Stecker (2002) for alternative 
detailed model calculations. The parameter values for all subsequent models are 
given in the respective figure captions.

While the model shown in Fig.\ \ref{sed01} gives a satisfactory fit to the 
X-ray to $\gamma$-ray data, it under-predicts the radio and optical fluxes. 
The model thus suggests that the low-energy radio to optical radiation is 
dominated by emission from other regions than those that emit the bulk 
of the X-rays and $\gamma$-rays.
This finding is consistent with the fact that we found much less flux variability
in the radio and optical bands than in the X-ray and $\gamma$-ray bands.
The Inverse Compton SED corrected for extragalactic absorption peaks in our model
at 1.7~TeV. Between 100 GeV and 400 GeV the CIB/COB model predicts a characteristic sharp 
turnover. The next generation Cherenkov telescopes CANGAROO~III, H.E.S.S., MAGIC, and 
Whipple should be able to measure such sharp turnovers in blazar energy spectra.

\begin{figure}[bh]
\begin{center}
\resizebox{10cm}{!}{
\plotone{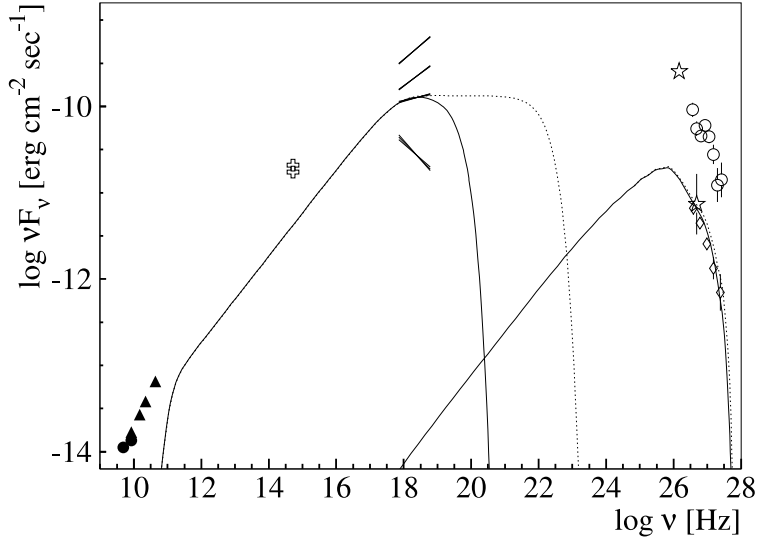}}
\end{center}
%\parbox[b]{55mm}{
\caption{\label{orphan1} SSC models of the data from the orphan $\gamma$-ray flare on
June 4, 2002. 
In addition to the data from Fig.\ \ref{sed01}, the open stars 
show the TeV flux estimates from 5 hrs before and during the orphan $\gamma$-ray flare 
(the X-ray flux stayed at a constant level during the flare).
The model fits the pre-flare and flare X-ray data, but only the pre-flare
$\gamma$-ray data. 
The two models computed with different high-energy cutoff of accelerated particles 
show that the additional highest energy electrons mainly produce Inverse Compton
emission at energies above those sampled by the observations (above $\sim$10~TeV).
In the model, the turn over of the $\gamma$-ray component originates 
from extragalactic extinction rather than from the high-energy cutoff 
of the electron energy spectrum.
The model parameters are: $\delta_{\rm j}\,=$ 20, $B\,=$ 0.04~G, 
$R\,=$ 1.4$\times$10$^{16}$~cm, log$(E_{\rm min}/\rm eV)\,=$ 3.5, 
log$(E_{\rm b}/\rm eV)\,=$ 11.45, $p_1\,=$ 2, $p_2\,=$ 3, 
electron energy density 0.014 erg cm$^{-3}$.
Solid line: log$(E_{\rm max}/\rm eV)\,=$ 12.2, dotted line:
log$(E_{\rm max}/\rm eV)\,=$ 13.5.
All models  include the effect of extragalactic absorption.
}
\end{figure}

We explored several ways to produce the orphan $\gamma$-ray flare in the framework of SSC models.
Given the observed {\it RXTE} energy spectrum and our choice of model parameters, 
it is not possible to produce an orphan $\gamma$-ray 
flare by moving the high-energy cutoff of accelerated electrons to higher energies 
(Fig.\ \ref{orphan1}).
The reason for this behavior is that high-energy electrons that emit synchrotron radiation
above the {\it RXTE} energy range, emit Inverse Compton $\gamma$-rays at energies above those 
sampled by the observations (above $\sim$10~TeV).
The additional photons show up at energies above $\sim$10 TeV. 
Extragalactic extinction reduces the flux above 10 TeV already by so much that
it is not shown in the figure.

\begin{figure}[bh]
\begin{center}
\begin{minipage}{8cm}
\resizebox{8cm}{!}{
\plotone{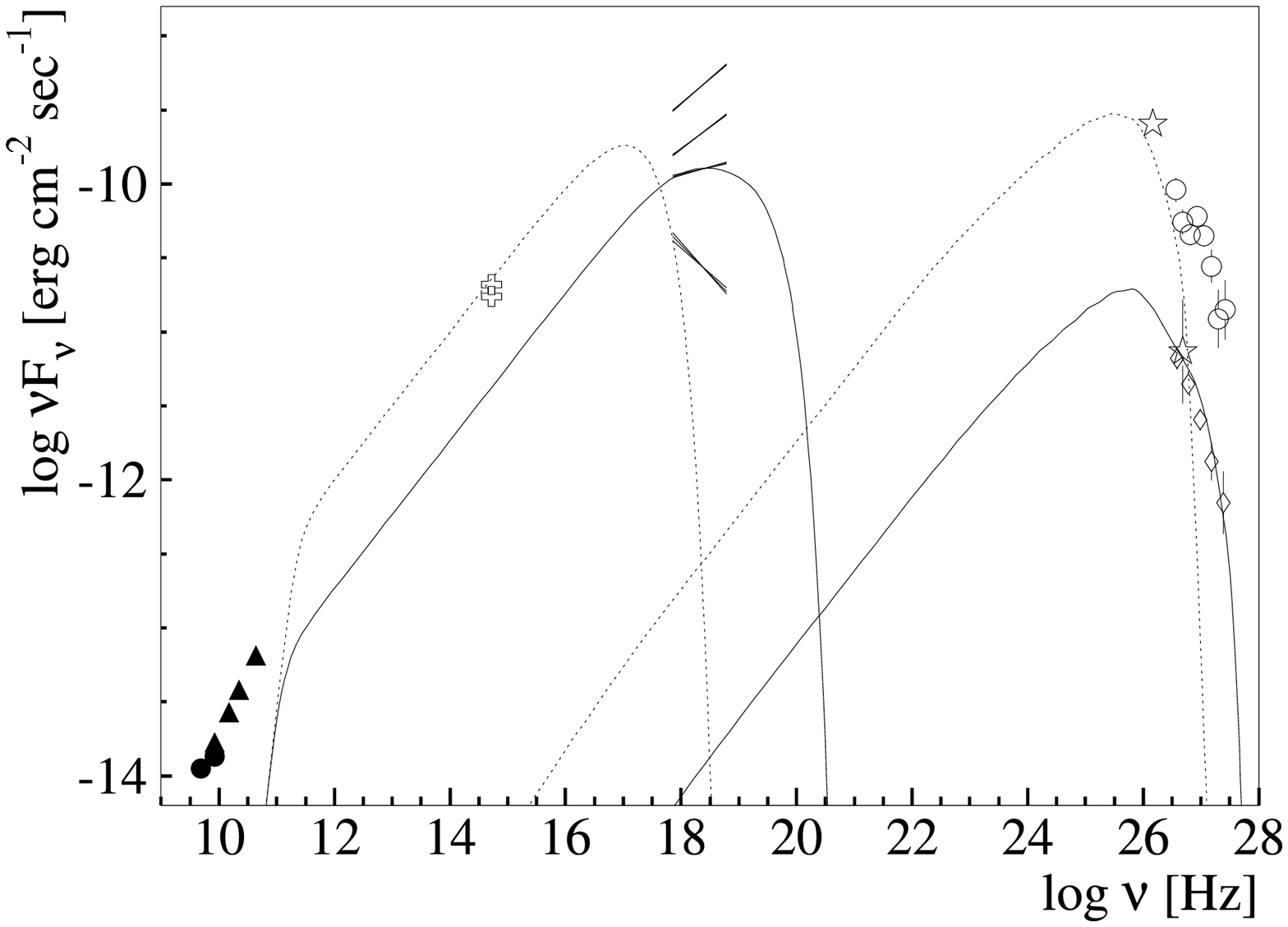}}
\end{minipage}
\begin{minipage}{8cm}
\resizebox{8cm}{!}{
\plotone{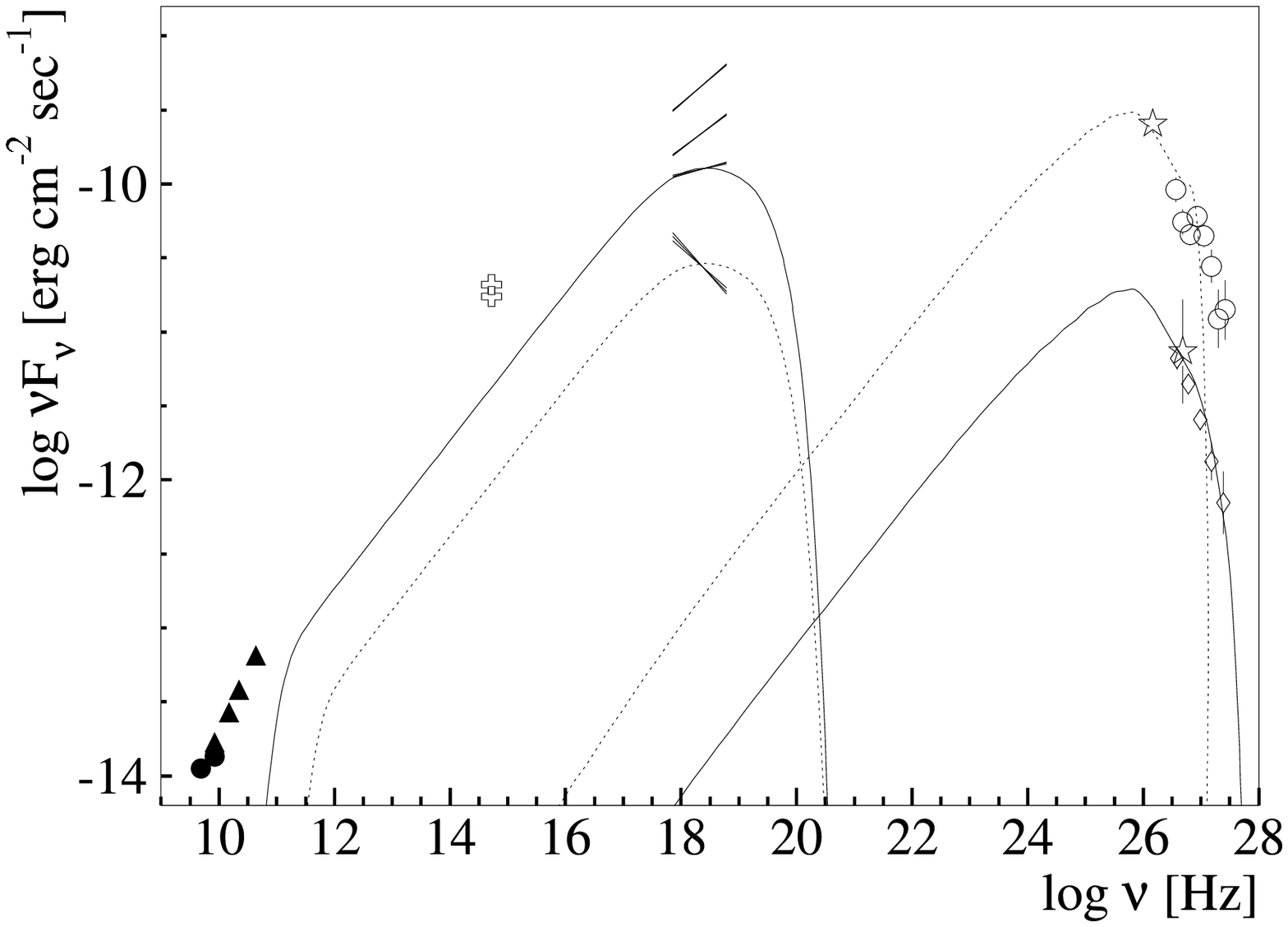}}
\end{minipage}
\end{center}
\caption{\label{orphan2} Same data as in Fig.\ \ref{orphan1}.
In both panels, the solid lines show the SSC model that explains the ``pre-flare'' X-ray and
$\gamma$-ray emission, and the dotted lines show additional emission during the
$\gamma$-ray flare. All models include the effect of extragalactic absorption.
In the left panel, the $\gamma$-ray flare is produced by an electron population 
with a rather low high-energy cutoff, log$(E_{\rm max}/\rm eV)\,=$ 11.15 instead of
log$(E_{\rm max}/\rm eV)\,=$ 12.2.
In the right panel, a dense electron population confined to a small emission region 
produces the orphan flare.
The model parameters for the flare component are as follows, left panel: 
$\delta_{\rm j}\,=$ 20, $B\,=$ 0.04~G, 
$R\,=$ 1.4$\times$10$^{16}$~cm, single electron power-law with 
log$(E_{\rm min}/\rm eV)\,=$ 3.5, log$(E_{\rm b}/\rm eV)\,=$ log$(E_{\rm max}/\rm eV)\,=$ 11.15, 
$p_1\,=$ 2, electron energy density 0.07 erg cm$^{-3}$;
right panel:
$\delta_{\rm j}\,=$ 20, $B\,=$ 0.04~G, 
$R\,=$ 8$\times$10$^{14}$~cm, log$(E_{\rm min}/\rm eV)\,=$ 3.5, 
log$(E_{\rm max}/\rm eV)\,=$ 12.2,  log$(E_{\rm b}/\rm eV)\,=$ 11.45,
$p_1\,=$ 2, $p_2\,=$ 3, electron energy density 17 erg cm$^{-3}$.
The parameters for the quiescent emission are the same as in Fig.\ \ref{orphan1}.
}
\end{figure}

Adding a low-energy electron population (Fig.\ \ref{orphan2}, left panel) succeeds in
producing an orphan $\gamma$-ray flare and predicts an extremely steep $\gamma$-ray 
energy spectrum.
However, the model needs careful fine-tuning, as the density of low-energy electrons 
is constrained by the optical measurements. 
Studies of the TeV $\gamma$-ray energy spectrum during the flare 
are underway to test the prediction of a steep spectrum.
A more natural way to explain the flare is to postulate a second, dense electron population 
within a small emission region (Fig.\ \ref{orphan2}, right panel); 
compared to the region where the quiescent emission comes from, 
the 1200 times larger energy density of this electron population and the 
5400 times smaller emission volume lead to a high Inverse Compton to 
synchrotron luminosity ratio, and thus to a $\gamma$-ray flare 
without a strong X-ray flare.
Note that this model does not suffer from a ``Compton catastrophe''.
The optical thickness for internal absorption in 
$\gamma_{\rm TeV}+\gamma_{\rm Seed}\,\rightarrow$ $e^+\,e^-$
pair-production processes is well below 1 over the full range of 
gamma-ray energies covered by the TeV observations.

\begin{figure}[bh]
\begin{center}
\resizebox{10cm}{!}{
\plotone{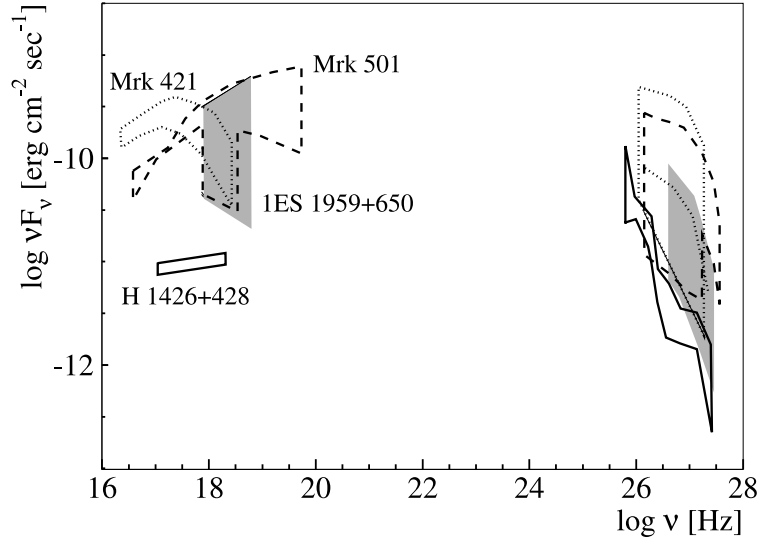}}
\end{center}
\caption{
Comparison of the ranges of X-ray and $\gamma$-ray energy spectra observed from
Mrk 421 (dotted line), Mrk 501 (dashed line), 1ES~1959+650 (shaded area), and H~1426+428
(solid line). 
For Mrk 421, BeppoSAX X-ray data from Fossati et al.\ (2000) and Whipple and HEGRA
$\gamma$-ray data from Krennrich et al.\ (2002) and Aharonian et al.\ (1999c) have been used.
For Mrk 501, BeppoSAX data from Pian et al.\ (1998), 
{\it RXTE} data from Krawczynski et al.\ (2000), and HEGRA $\gamma$-ray data from 
Aharonian et al.\ (1999b,2001) entered the graphs;
the different energy coverages of the BeppoSAX ($\sim$0.15-150~keV) and {\it RXTE} (3-25 keV) 
satellites resulted in the complex shape of the region of observed X-ray fluxes.
For H~1426+428 the X-ray data are from (Giommi et al.\ 2002). 
and the $\gamma$-ray data are from \cite{Petr:02,Djan:02,Ahar:03a}.
The 1ES~1959+650 the X-ray data are from this work and the $\gamma$-ray 
data are from \cite{Ahar:03b}.
\label{sed02}}
\end{figure}

In Fig.\ \ref{sed02} we compare the X-ray and $\gamma$-ray energy spectra
of 1ES~1959+650 with those of the three other TeV blazars with measured 
TeV energy spectra.
The X-ray and $\gamma$-ray energy spectra of 1ES 1959+650 are very similar
to those of Mrk 501. In comparison to these two sources, the X-ray spectra
of Mrk 421 are softer, while the $\gamma$-ray energy spectra are similar.
The X-ray energy spectrum of H1426+428 seems to be relatively hard.
A meaningful comparison of the high-energy TeV $\gamma$-ray energy spectra of H~1426+428 and
the other 3 sources is hampered by the highly uncertain extent of 
extragalactic absorption for the high-redshift source H~1426+428.
\section{Correlation Between Emission Parameters and Black Hole Mass Indicators}
\label{bhm}
With the addition of another TeV blazar with good broadband data,
we consider the set of all TeV blazars to begin to look for a connection
of the jet properties to the properties of the central accreting black hole
thought to drive the jet.
Ferrarese \& Merrit (2000) and Gebhardt (2000) discovered a close correlation between
the mass of the central black holes, $M_{\bullet}$, and the host galaxy's 
stellar velocity dispersion, $\sigma_*$. 
The present data on nearby galaxies do not show evidence for an intrinsic scatter 
of the correlation and the upper limit on the width of the correlation is $0.4\,M_{\bullet}$. 
The correlation is significantly tighter than that of $M_{\bullet}$ and
the galactic bulge luminosity, $L_{\rm blg}$.
Based on both correlations Falomo, Kotilainen \& Treves (2002) and 
Barth et al. (2003) estimated the black hole masses of several BL Lac objects, 
including 5 of the 6 established TeV-blazars. 
For 1ES~1959+650, Falomo et al.\ estimated $\log{(M_\bullet/M_\odot)}\,=$ 8.12$\pm$0.13,
using the $M_{\bullet}$-$\sigma_*$ correlation.
As can be seen from the black hole masses given in Table~\ref{blazars}, the black hole of 
1ES~1959+650 seems to be the least massive of all TeV blazars and is separated by 
one order of magnitude from the most massive one, Mrk~501.

\begin{figure}[bh]
\begin{center}
\resizebox{13cm}{!}{
\epsscale{0.8}
\plotone{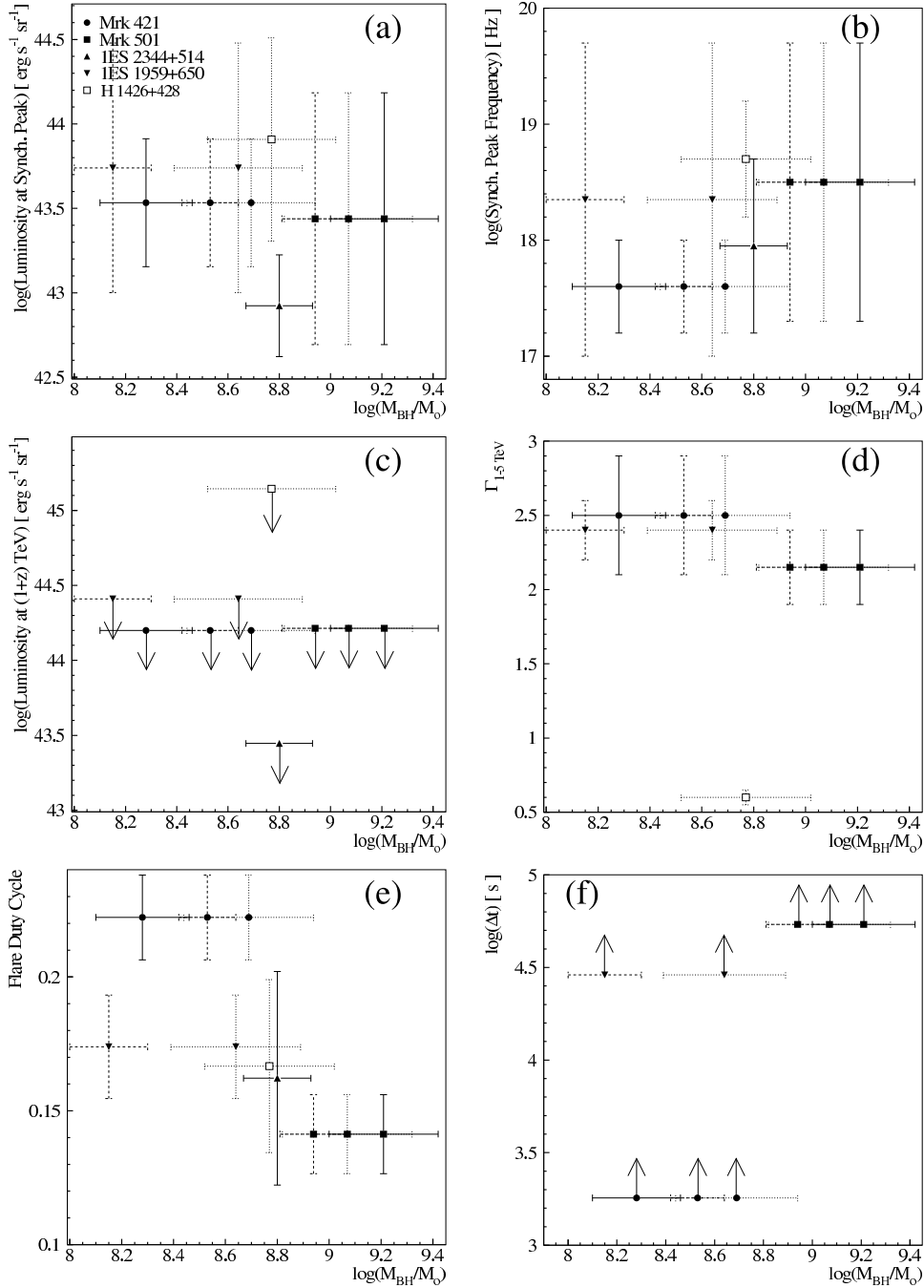}}
\end{center}
%\parbox[b]{55mm}{
\caption{\scriptsize \label{mb1} The correlation between the black hole mass estimator and various
parameters describing the characteristics of the X-ray and $\gamma$-ray emission for the five
well-established TeV blazars for which black hole masses have been estimated:
(a) the peak luminosity of the low-energy component; 
(b) frequency at which the low-energy SED peaks; 
(c) the range of observed $\gamma$-ray luminosities at (1+$z$) TeV; 
(d) the 1-5 TeV photon index;
(e) the flare duty cycle $f_{\rm flr}$ determined from 2-12 keV {\it RXTE} ASM data, and
(f) the range of exponential increase/decay constants observed at $\gamma$-ray energies.
In all panels, the symbols identify the blazars according to the legend given in the
first panel; black hole masses from stellar velocity dispersion measurements
are shown with solid (Barth et al.\ 2003) and dashed (Falomo et al.\ 2002) error bars; 
black hole masses from galactic bulge luminosities (Falomo et al.) 
are shown with dotted error bars.
Horizontal error bars are statistical errors in the case of solid and dashed lines, 
for the dotted estimates no statistical errors have been published and we assumed 
$\Delta log(M_\bullet)\,=$ 0.25.
The vertical error bars show the ranges of observed values.
References are given in Table~\ref{blazars}; the fastest $\gamma$-ray $e$-folding times
for Mrk 421, Mrk 501, and 1ES 1959+650 are from Gaidos et al.\ (1996), 
Aharonian et al.\ (1999a), and Holder et al.\ (2002), respectively.
The available data limit panel (d) to 4 sources and panel (f) to 3 sources.
}
\end{figure}

The black hole mass estimates allow us to explore the correlation between mass 
and parameters describing the jet emission and therefore with the jet properties.
In Fig.\ \ref{mb1}a-f we show the correlation between the black hole mass and 
6 parameters that characterize the jet continuum emission:
(i) the luminosity at the peak of the low-energy (synchrotron) emission component;
(ii) the frequency at which the low-energy SED peaks;
(iii) the range of observed luminosities at (1+$z$)~TeV;
(iv) the 1--5 TeV photon index;
(v) the X-ray ``flare duty cycle'', $f_{\rm flr}$, defined
as the fraction of time during which the {\it RXTE} ASM flux exceeds 50\% 
of the time averaged flux (see more detailed description below); and,
(vi) the range of $\gamma$-ray $e$-folding time observed so far.
The first two parameters describe the SED of the low-energy (synchrotron) component;
the second two parameters describe the SED of the high-energy (Inverse Compton) component;
the last two parameters describe temporal properties of the X-ray and the $\gamma$-ray
emission. 

The $\gamma$-ray parameters were corrected for extragalactic extinction 
based on the CIB/COB model of Kneiske et al.\ (2002). The flare duty cycle was computed from the
{\it RXTE} ASM data taken between 1996 and mid 2003. Binning the data into 28-day bins, 
we determined the fraction of bins where the flux surpassed the mean flux from 
that source by 50\%. We determined error bars on these duty cycles with a Monte-Carlo
simulation, by re-calculating the values 1000 times, modifying each time the 
flux values according to a Gaussian distribution with a width given by the
experimental flux errors.
The choice of the time binning changes the values of the flare duty cycle,
but does not change the results qualitatively.
The symbols differentiate the sources; the solid and dashed error bars show the black hole 
mass estimates based on stellar velocity dispersion measurements from 
Falomo et al.\ (2002) and Barth et al.\ (2003), respectively.
The dotted error bars show the black hole mass estimates from bulge 
luminosity measurements (also from Falomo et al.\ 2002).
Horizontal error bars show the statistical uncertainty on the $M_\bullet$-estimates, and
the vertical error bars show the ranges of observed values. 
Differences between parameter ranges can be highly significant 
from a statistical point of view even if the vertical ``error bars'' span various 
orders of magnitudes and exhibit a substantial overlap.
The figures do not show clear correlations. The only quantity that shows an indication 
for a correlation is the X-ray flare duty cycle $f_{\rm flr}$.
\section{Discussion}
\label{disc}
Early SSC modeling of Mrk 421 and Mrk 501 data indicated that simple 
one-zone SSC models were capable of describing a wealth of data satisfactorily
(Inoue \& Takahara 1996, Takahashi et al.\ 2000, Krawczynski et al.\ 2001).
For Mrk 501 however, detailed time dependent modeling showed that the very simplest
SSC models failed to account for the combined broadband X-ray (BeppoSAX, {\it RXTE})
and TeV $\gamma$-ray data \cite{Kraw:02}. In order to consistently fit the data 
from several flares, 
the authors had to introduce a second emission zone, as well as a 
poorly justified ``minimum Lorentz factor of accelerated electrons'' on the order 
of $\gamma_{\rm min}\,=$ 10$^5$ and higher.
In this paper we presented evidence for an ``orphan'' $\gamma$-ray flare 
without X-ray counterpart. Also this finding contradicts the most simple
1-zone SSC models. There are several ways to explain the orphan flare:
\begin{itemize}
\item {\bf Multiple-Component SSC Models:} 
A high density electron population confined to a small emission volume 
can account for an orphan $\gamma$-ray flare (see Sect.\ \ref{sed}).
``Low duty-cycle'' fast variability has been observed for a number of sources. 
A prime example is the detection of a strong X-ray flare from Mrk 501 
with a doubling time of 6 minutes \cite{Cata:00}. Such observations strongly 
suggest that indeed small regions with high electron densities produce 
strong and rapid flares. Alternatively, a second electron population 
with a low high-energy cutoff might produce an orphan $\gamma$-ray flare, 
as mentioned above. However, the corresponding SSC model requires 
fine-tuning of the model parameters. As a third possibility a population 
of electrons with a very hard energy spectrum might produce the 
gamma-ray flare while it emits synchrotron radiation at energies above 
those sampled with the {\it RXTE}.
\item {\bf External Compton Models:} In External Compton models, the $\gamma$-ray flux 
originates from Inverse Compton processes of high-energy electrons with radiation 
external to the jet. Variations of the external photon intensity in the jet frame  
can cause $\gamma$-ray flares without lower-energy counterparts.
Such variations could have different origins: the external photon flux, e.g.\ from
the accretion disk, could be intrinsically variable.
Alternatively, the motion of the emission region relative to an external photon
reflector could result in a time-variable photon 
flux in the jet frame \cite{Wehr:98}.
In External Compton models, the external photon field is highly anisotropic in the jet frame,
owing to the highly relativistic motion of the jet plasma ($\Gamma\gg10$).
As a consequence, the Inverse Compton emission has a narrower
beaming angle than the synchrotron emission and 
a slight precession of the jet could cause a large change in the TeV flux accompanied by
a small change of the X-ray flux.
\item {\bf Magnetic Field Aligned along Jet Axis:}
If the magnetic field in the emission region of the orphan flare 
is aligned with the jet axis and thus with the line of sight, 
the observer would not see the synchrotron flare. 
The electrons however would scatter SSC gamma-rays into our 
direction and we would thus be able to see the Inverse Compton flare.
\item {\bf Proton Models:} 
In proton models the low-energy radiation is produced by a population of non-thermal 
electrons and high-energy radiation by accelerated protons, 
either directly as synchrotron radiation \cite{Ahar:00a,Muec:02}, or via a 
Proton Induced Cascade (PIC) \cite{Mann:98}. 
As electron and proton injection rates and high-energy cutoffs may vary in a different
way with the plasma conditions, proton models naturally account for orphan flares.
In PIC models, the TeV $\gamma$-ray emission originates from a thin surface layer 
of an optically thick pair plasma, while the X-ray emission originates from 
the full emission volume. 
The model naturally accounts for orphan $\gamma$-ray flares, as the thin surface 
layer can produce more rapid flares than the larger X-ray emission region.
We consider it unlikely that this latter explanation applies to the observation 
of the orphan flare from 1ES~1959+650, as the X-ray and $\gamma$-ray fluxes 
varied on comparable time scales throughout the rest of the observation campaign.
\end{itemize}
Our main conclusion from the observation of the orphan $\gamma$-ray 
flare is that it can not be explained with conventional one-zone SSC models.

The black hole mass estimates from stellar velocity dispersion measurements allow us
to study the connection between the jet emission parameters and the central black hole mass.
We expect to find correlations as the characteristic length and time scales
of the accretion system scale with $M_\bullet$ 
(see e.g., Mirabel et al.\ 1992). Our data however, did not reveal any correlations. 
It is remarkable that 1ES~1959+650 and Mrk~501 show very similar X-ray and $\gamma$-ray 
energy spectra and flux variation time scales while their black hole masses 
differ by about one order of magnitude.

Variations of parameters like jet viewing angle, jet magnetic field, or 
intensity and energy spectrum of the ambient photon field may mask the correlations.
Furthermore, our correlation plots suffer from the limitations of the observations:
flux threshold selection effects, limited energy coverage of the observations, and
short time over which the data were acquired (relative to the lifetime of the jet).
Alternatively, the $M_\bullet$--$\sigma_*$ and $M_\bullet$--$L_{\rm blg}$ correlations
found for nearby galaxies may not hold for blazars, rendering the black hole
mass estimates used in our analysis inaccurate (Barth et al.\ 2003).
\hspace*{2cm}\\[2ex]
{\it Acknowledgements:}
We thank Jean Swank, David Smith and the {\it RXTE} GOF for their
excellent collaboration in scheduling the {\it RXTE} observations.
We thank the VERITAS and HEGRA collaborations for the TeV $\gamma$-ray
light curves and energy spectra. HK and SH gratefully acknowledge support by NASA 
through the grant NASA NAG5-12974. AM acknowledges support by
the National Science Foundation grant AST-0098579.
The University of Michigan Radio Astrophysical Observatory (UMRAO) 
is operated by funds from the University of Michigan Department 
of Astronomy. We acknowledge helpful comments by an anonymous referee.
%

%\onecolumn
\clearpage
%
% {\it RXTE} ASM
%


%
% X/TeV correlation
%
%
% Flux-Hardness correlation
%

%
% SEDs
%
%
% Possibilities
%
%
% Comp.
%
%
% M_BH - Correlations #1
%

%
%
%
%\begin{figure}[bh]
%\begin{center}
%\resizebox{8cm}{!}{\plotone{m05.epsi}}
%\end{center}
%%\parbox[b]{55mm}{
%\caption{\label{deltagamma} Steepening of the 1-5 TeV $\gamma$-ray photon index
%$\Gamma$ by extragalactic $\gamma$-ray absorption as function of redshift.
%A $\gamma$-ray source at redshift $z$ that would produce an energy spectrum of 
%photon index $\Gamma_0$ at $z\,=$ 0 if no extragalactic absorption existed, 
%will indeed produce an energy spectrum of photon index $\Gamma\,=$ $\Gamma_0\,+$ 
%$\Delta\Gamma$. The calculation uses the COIB model of Kneiske et al.\ (2002).}
%\end{figure}
%
% table of TeV candidates
%
\clearpage

\setcounter{page}{30}

\setlength{\textwidth}{20cm}
\hspace*{-2cm}
\begin{deluxetable}{p{1.6cm}p{0.7cm}p{2cm}p{1.5cm}p{1.5cm}p{1.5cm}p{1.3cm}p{1cm}c}
%\rotate
\tabletypesize{\footnotesize}
\tablewidth{0pt}
\tablecaption{
Properties of Established TeV-Blazars. \label{blazars}\newline
References: 
($r_1$) Giommi et al.\ (2002); 
($r_2$) Fossati et al.\ (2000); 
($r_3$) Gaidos et al.\ (1996);\newline
($r_4$) Krennrich et al.\ (2002);
($r_5$) Aharonian et al.\ (1999c);
($r_6$) Barth et al.\ (2003);\newline
($r_7$) Falomo et al.\ (2002);
($r_8$) Pian et al.\ (1998);
($r_9$) Sambruna et al.\ (2000);\newline
($r_{10}$) Aharonian et al.\ (1999a);
($r_{11}$) Aharonian et al.\ (1999b,2001);
($r_{12}$) Catanese et al.\ (1998);\newline
($r_{13}$) this work;
($r_{14}$) Holder et al.\ (2002);
($r_{15}$) Aharonian et al.\ (2003b);\newline
($r_{16}$) Chadwick et al.\ (1999);
($r_{17}$) Horan et al.\     (2002);
($r_{18}$) Aharonian et al.\ (2003a).}
\tablehead{
\colhead{Object} & 
\colhead{$z$} & 
\colhead{References for} &
\colhead{log$\,\nu_{\rm S}$ $^a$}&
\colhead{log$\,L_{\rm x}$ $^b$}&
\colhead{log$\,L_{\rm \gamma}$ $^c$}&
\colhead{$\Gamma_{\rm \gamma}$ $^d$}&
\colhead{log $^e$}&
\colhead{log $^f$}\\
&&\colhead{TeV Detections}&
\colhead{$\left[\rm Hz\right]$}&
\colhead{$\rm \left[erg\, s^{-1}\,sr^{-1}\right]$}&
\colhead{$\rm \left[erg\, s^{-1}\,sr^{-1}\right]$}&&
\colhead{M$_\sigma$/M$_\odot$} & 
\colhead{M$_{\rm blg}$/M$_\odot$}
}
\startdata
 Mrk 421     & 0.031 &  Punch et al.\     1992, Petry et al.\     1996 &
17.2--18.0$^{r_1,r_2}$  & 43.2--43.9$^{r_1,r_2}$ & 44.2 (450)$^{r_3}$   & 2.1--2.9 (2.5--3.3)$^{r_4,r_5}$ 
&8.28$\pm$0.11$^{r_6}$ 8.50$\pm$0.18$^{r_7}$ & 8.69$^{r_7}$ \\
 Mrk 501     & 0.034 &  Quinn et al.\     1996, Bradbury et al.\  1997  &
17.3--19.7$^{r_8,r_9}$   & 42.7--44.2$^{r_8,r_9}$ & 44.2 (365)$^{r_{10}}$ & 1.9--2.4 (2.3--2.8)$^{r_{10},r_{11}}$ 
&9.21$\pm$0.13$^{r_6}$ 8.93$\pm$0.21$^{r_7}$ & 9.07$^{r_7}$ \\
1ES~2344+514 & 0.044 &  Catanese et al.\  1998, Tluczykont et al.\ 2003 &
17.2--18.7$^{r_1}$   & 42.6--43.2$^{r_1}$ & 43.4 (30)$^{r_{12}}$    & ---        &8.80$\pm$0.16$^{r_6}$ & ---  \\
1ES~1959+650 & 0.047 &  Nishiyama et al.\ 1999, Holder et al.\ 2003,
Aharonian et al.\ 2003b &
17.0--19.7$^{r_1,r_{13}}$ & 43.0--44.5$^{r_1,r_{13}}$ & 44.4 (215)$^{r_{14}}$   & 2.2--2.6 (2.8-3.2)$^{r_{15}}$ 
&8.12$\pm$0.13$^{r_7}$ & 8.64$^{r_7}$ \\
PKS~2155-304 & 0.116 &  Chadwick et al.\  1999, Hinton et al.\    2003 &
16.0--17.0$^{r_1}$   & 44.4--44.8$^{r_1}$ & 45.1 (45)$^{r_{16}}$    & ---        &--- & --- \\
  H~1426+428 & 0.129 &  Horan et al.\     2002, Aharonian et al.\ 2003a&
18.2--19.2$^{r_1}$   & 43.3--44.5$^{r_1}$ & 45.1 (30)$^{r_{17},r_{18}}$    &  0.6 (2.2)$^{r_{18}}$ &--- & 8.77$^{r_7}$ \\
\enddata
\tablenotetext{a}{\hspace*{0.2cm} Frequency range over which synchrotron peak has been detected.}
\tablenotetext{b}{\hspace*{0.2cm} Range of observed synchrotron peak luminosities.}
\tablenotetext{c}{\hspace*{0.2cm} Highest (1+$z$) TeV luminosities observed, including
the correction for extragalactic extinction according to Kneiske et al.\ 2002. 
The number in parenthesis gives the corresponding flux at 1 TeV in 
(10$^{-12}$ erg cm$^{-2}$ s$^{-1}$).}
\tablenotetext{d}{\hspace*{0.2cm} Range of observed 1-5 TeV photon indices, corrected for 
extragalactic extinction. Numbers in parenthesis give the photon indices before correction.}
\tablenotetext{e}{\hspace*{0.2cm} Mass of central black hole from stellar velocity dispersion.}
\tablenotetext{f}{\hspace*{0.2cm} Mass of central black hole from bulge luminosity.}
\end{deluxetable}
%
% X-ray power-law fits 
%
\begin{deluxetable}{cccccc}
\tablecaption{Results of power-law fits to the 3 keV -- 25 keV data\newline
(Statistical errors only) \label{xflux}}
\tablewidth{0pt}
\tabletypesize{\footnotesize}
\tablehead{
\colhead{Start MJD} & 
\colhead{$t_{\rm obs}\,$\tablenotemark{a}} & 
\colhead{$F_{\rm 10\,keV}\,$\tablenotemark{b}} & 
\colhead{$\Gamma\,$\tablenotemark{c}} &
\colhead{$\chi^2_{\rm r} \,/\,\rm d.o.f.\,$\tablenotemark{d}} &
\colhead{$P_{\rm c}\,$\tablenotemark{e}}
}
\startdata
    52412.1406 & 0.45 & 1.550 $\pm$ 0.011 & 1.831 $\pm$ 0.010 &  0.68 / 45&    0.95\\
    52412.3477 & 0.36 & 2.229 $\pm$ 0.014 & 1.772 $\pm$ 0.009 &  0.61 / 45&    0.98\\
    52412.9922 & 0.16 & 1.028 $\pm$ 0.014 & 2.006 $\pm$ 0.018 &  1.46 / 45&    0.02\\
    52413.1406 & 0.12 & 1.865 $\pm$ 0.023 & 1.717 $\pm$ 0.018 &  0.53 / 45&    1.00\\
    52413.3516 & 0.24 & 2.277 $\pm$ 0.018 & 1.710 $\pm$ 0.011 &  0.67 / 45&    0.96\\
    52413.7734 & 0.43 & 2.285 $\pm$ 0.011 & 1.717 $\pm$ 0.007 &  0.93 / 45&    0.60\\
    52414.4609 & 0.40 & 2.937 $\pm$ 0.016 & 1.660 $\pm$ 0.008 &  1.28 / 45&    0.10\\
    52415.1367 & 0.14 & 1.292 $\pm$ 0.017 & 1.878 $\pm$ 0.019 &  0.82 / 45&    0.80\\
    52415.3828 & 0.42 & 1.202 $\pm$ 0.010 & 1.862 $\pm$ 0.011 &  0.51 / 45&    1.00\\
    52416.1211 & 0.16 & 1.151 $\pm$ 0.015 & 1.891 $\pm$ 0.018 &  0.70 / 45&    0.93\\
    52416.4375 & 0.43 & 1.183 $\pm$ 0.010 & 1.797 $\pm$ 0.012 &  0.59 / 45&    0.99\\
    52417.1758 & 0.23 & 0.824 $\pm$ 0.011 & 1.995 $\pm$ 0.018 &  0.45 / 45&    1.00\\
    52417.8672 & 0.24 & 1.155 $\pm$ 0.013 & 1.755 $\pm$ 0.016 &  0.89 / 45&    0.68\\
    52417.3594 & 0.43 & 0.986 $\pm$ 0.008 & 1.909 $\pm$ 0.010 &  0.85 / 45&    0.75\\
    52417.4258 & 0.44 & 0.892 $\pm$ 0.007 & 1.952 $\pm$ 0.010 &  0.49 / 45&    1.00\\
    52418.4141 & 0.44 & 1.485 $\pm$ 0.008 & 1.678 $\pm$ 0.008 &  0.54 / 45&    0.99\\
    52419.4062 & 0.17 & 0.927 $\pm$ 0.012 & 1.860 $\pm$ 0.017 &  0.67 / 45&    0.95\\
    52420.4648 & 0.33 & 1.339 $\pm$ 0.010 & 1.795 $\pm$ 0.010 &  0.56 / 45&    0.99\\
    52421.8477 & 0.08 & 1.166 $\pm$ 0.019 & 1.821 $\pm$ 0.023 &  0.95 / 45&    0.57\\
    52422.9102 & 0.16 & 1.142 $\pm$ 0.015 & 1.891 $\pm$ 0.016 &  0.76 / 45&    0.88\\
    52423.3672 & 0.24 & 0.995 $\pm$ 0.010 & 1.867 $\pm$ 0.013 &  0.75 / 45&    0.89\\
    52424.7891 & 0.19 & 0.801 $\pm$ 0.011 & 2.001 $\pm$ 0.017 &  0.31 / 45&    1.00\\
    52425.2188 & 0.17 & 0.658 $\pm$ 0.011 & 2.041 $\pm$ 0.023 &  0.90 / 45&    0.67\\
    52426.2070 & 0.17 & 1.179 $\pm$ 0.015 & 1.860 $\pm$ 0.018 &  0.64 / 45&    0.97\\
    52427.3047 & 0.46 & 1.147 $\pm$ 0.009 & 1.817 $\pm$ 0.011 &  1.02 / 45&    0.43\\
    52428.3047 & 0.47 & 1.156 $\pm$ 0.009 & 1.856 $\pm$ 0.011 &  0.85 / 45&    0.75\\
    52429.0859 & 0.33 & 0.793 $\pm$ 0.009 & 1.874 $\pm$ 0.016 &  0.77 / 45&    0.86\\
    52429.3008 & 0.27 & 0.793 $\pm$ 0.011 & 1.854 $\pm$ 0.018 &  0.58 / 45&    0.99\\
    52429.3398 & 0.65 & 0.750 $\pm$ 0.007 & 1.913 $\pm$ 0.012 &  1.06 / 45&    0.36\\
    52429.6016 & 0.17 & 0.503 $\pm$ 0.010 & 2.051 $\pm$ 0.026 &  0.92 / 45&    0.63\\
    52430.3477 & 0.49 & 0.577 $\pm$ 0.007 & 1.928 $\pm$ 0.016 &  0.53 / 45&    1.00\\
    52430.2969 & 0.10 & 0.563 $\pm$ 0.014 & 2.033 $\pm$ 0.033 &  0.72 / 45&    0.92\\
    52431.0820 & 0.16 & 0.751 $\pm$ 0.013 & 1.931 $\pm$ 0.023 &  0.49 / 45&    1.00\\
    52431.3164 & 0.93 & 0.789 $\pm$ 0.007 & 1.773 $\pm$ 0.012 &  1.15 / 45&    0.22\\
    52432.3047 & 0.61 & 0.498 $\pm$ 0.007 & 2.041 $\pm$ 0.018 &  0.74 / 45&    0.91\\
    52432.1172 & 0.75 & 0.654 $\pm$ 0.006 & 1.938 $\pm$ 0.012 &  0.46 / 45&    1.00\\
    52432.8359 & 0.26 & 0.521 $\pm$ 0.009 & 1.953 $\pm$ 0.023 &  0.76 / 45&    0.88\\
    52433.9141 & 0.28 & 0.371 $\pm$ 0.007 & 2.119 $\pm$ 0.024 &  0.64 / 45&    0.97\\
    52433.2930 & 2.52 & 0.583 $\pm$ 0.005 & 1.937 $\pm$ 0.011 &  0.52 / 45&    1.00\\
    52434.1562 & 0.85 & 0.477 $\pm$ 0.005 & 2.017 $\pm$ 0.013 &  0.44 / 45&    1.00\\
    52434.3477 & 0.94 & 0.399 $\pm$ 0.003 & 2.087 $\pm$ 0.011 &  1.39 / 45&    0.04\\
    52434.6797 & 0.24 & 0.489 $\pm$ 0.009 & 2.040 $\pm$ 0.023 &  0.72 / 45&    0.92\\
    52435.3398 & 0.94 & 0.430 $\pm$ 0.005 & 2.045 $\pm$ 0.016 &  0.80 / 45&    0.82\\
    52436.3281 & 0.72 & 0.428 $\pm$ 0.006 & 2.105 $\pm$ 0.020 &  1.29 / 45&    0.09\\
    52437.3164 & 0.70 & 0.291 $\pm$ 0.005 & 2.281 $\pm$ 0.024 &  0.99 / 45&    0.50\\
    52438.3047 & 0.68 & 0.226 $\pm$ 0.005 & 2.169 $\pm$ 0.029 &  0.64 / 45&    0.97\\
    52439.0938 & 0.54 & 0.180 $\pm$ 0.004 & 2.345 $\pm$ 0.025 &  0.91 / 45&    0.64\\
    52439.3008 & 0.47 & 0.182 $\pm$ 0.006 & 2.340 $\pm$ 0.041 &  0.62 / 45&    0.98\\
    52439.3672 & 0.47 & 0.170 $\pm$ 0.007 & 2.384 $\pm$ 0.051 &  0.93 / 45&    0.61\\
    52440.0859 & 0.95 & 0.248 $\pm$ 0.003 & 2.183 $\pm$ 0.017 &  0.81 / 45&    0.81\\
    52440.2891 & 0.48 & 0.384 $\pm$ 0.009 & 2.058 $\pm$ 0.030 &  0.73 / 45&    0.91\\
    52440.3516 & 0.56 & 0.426 $\pm$ 0.008 & 2.002 $\pm$ 0.026 &  0.60 / 45&    0.98\\
    52440.4219 & 0.60 & 0.490 $\pm$ 0.007 & 1.942 $\pm$ 0.021 &  1.19 / 45&    0.18\\
    52441.0742 & 2.53 & 0.221 $\pm$ 0.003 & 2.200 $\pm$ 0.015 &  0.79 / 45&    0.84\\
    52441.3359 & 0.30 & 0.219 $\pm$ 0.006 & 2.107 $\pm$ 0.034 &  0.56 / 45&    0.99\\
    52442.1289 & 0.70 & 0.157 $\pm$ 0.003 & 2.335 $\pm$ 0.022 &  0.77 / 45&    0.87\\
    52442.3320 & 0.54 & 0.196 $\pm$ 0.006 & 2.251 $\pm$ 0.042 &  0.98 / 45&    0.51\\
    52442.3984 & 0.58 & 0.212 $\pm$ 0.005 & 2.218 $\pm$ 0.029 &  0.91 / 45&    0.65\\
    52443.0781 & 1.91 & 0.239 $\pm$ 0.003 & 2.148 $\pm$ 0.016 &  0.84 / 45&    0.77\\
    52443.3789 & 0.72 & 0.248 $\pm$ 0.004 & 2.142 $\pm$ 0.020 &  0.95 / 45&    0.58\\
    52444.1055 & 0.96 & 0.479 $\pm$ 0.005 & 1.911 $\pm$ 0.015 &  0.75 / 45&    0.89\\
    52444.3750 & 0.56 & 0.333 $\pm$ 0.005 & 2.134 $\pm$ 0.022 &  0.80 / 45&    0.83\\
    52445.4219 & 0.47 & 0.249 $\pm$ 0.005 & 2.127 $\pm$ 0.024 &  0.94 / 45&    0.58\\
    52467.2266 & 0.39 & 0.961 $\pm$ 0.009 & 1.678 $\pm$ 0.015 &  0.60 / 45&    0.98\\
    52468.2148 & 0.44 & 0.598 $\pm$ 0.007 & 1.885 $\pm$ 0.017 &  0.51 / 45&    1.00\\
    52469.2031 & 0.33 & 1.249 $\pm$ 0.012 & 1.625 $\pm$ 0.014 &  0.68 / 45&    0.95\\
    52470.1406 & 0.20 & 1.883 $\pm$ 0.018 & 1.555 $\pm$ 0.014 &  0.60 / 45&    0.99\\
    52471.1797 & 0.25 & 1.028 $\pm$ 0.012 & 1.734 $\pm$ 0.017 &  1.45 / 45&    0.03\\
    52472.1680 & 0.25 & 1.012 $\pm$ 0.012 & 1.805 $\pm$ 0.017 &  0.68 / 45&    0.95\\
    52473.1562 & 0.28 & 0.963 $\pm$ 0.008 & 1.773 $\pm$ 0.012 &  0.81 / 45&    0.82\\
    52485.9297 & 0.54 & 1.182 $\pm$ 0.005 & 1.812 $\pm$ 0.007 &  0.99 / 45&    0.49\\
    52486.9219 & 0.56 & 0.989 $\pm$ 0.005 & 1.875 $\pm$ 0.008 &  0.98 / 45&    0.50\\
    52487.9688 & 0.76 & 1.089 $\pm$ 0.006 & 1.820 $\pm$ 0.007 &  0.62 / 45&    0.98\\
    52488.9570 & 0.93 & 1.094 $\pm$ 0.006 & 1.824 $\pm$ 0.008 &  0.82 / 45&    0.80\\
    52489.9453 & 0.94 & 0.925 $\pm$ 0.006 & 1.762 $\pm$ 0.009 &  1.09 / 45&    0.31\\
    52490.9336 & 0.94 & 1.226 $\pm$ 0.007 & 1.746 $\pm$ 0.008 &  0.87 / 45&    0.71\\
    52491.9414 & 0.46 & 1.415 $\pm$ 0.010 & 1.759 $\pm$ 0.010 &  0.48 / 45&    1.00\\
    52493.9180 & 0.50 & 1.003 $\pm$ 0.006 & 1.988 $\pm$ 0.008 &  0.87 / 45&    0.72\\
    52494.9727 & 0.50 & 1.017 $\pm$ 0.008 & 1.951 $\pm$ 0.011 &  0.72 / 45&    0.92\\
    52495.9609 & 0.53 & 1.079 $\pm$ 0.008 & 1.973 $\pm$ 0.010 &  0.79 / 45&    0.84\\
    52496.9297 & 0.51 & 1.362 $\pm$ 0.009 & 1.844 $\pm$ 0.010 &  0.82 / 45&    0.80\\
    52497.9297 & 0.60 & 1.185 $\pm$ 0.008 & 1.961 $\pm$ 0.009 &  0.76 / 45&    0.88\\
    52498.9727 & 0.56 & 1.167 $\pm$ 0.009 & 1.981 $\pm$ 0.011 &  0.72 / 45&    0.92\\
\enddata
\tablenotetext{a}{\hspace*{0.2cm} Duration in hours.}
\tablenotetext{b}{\hspace*{0.2cm} 10 keV flux in units of ($10^{-3}$ photons keV$^{-1}$ cm$^{-2}$ s$^{-1}$)}
\tablenotetext{c}{\hspace*{0.2cm} 3-25 keV photon index}
\tablenotetext{d}{\hspace*{0.2cm} Reduced $\chi^2$-value and degrees of freedom of the power-law fit}
\tablenotetext{e}{\hspace*{0.2cm} Chance probability for larger reduced $\chi^2$-values}
\end{deluxetable}
%
% 
%
%
% INCREASES
%
\begin{deluxetable}{cccc}

\tablecaption{Shortest e-folding times of 10 keV \newline
flux increases and decreases \label{inc}}
\tablewidth{0pt}
\tablehead{
\colhead{MJD1\tablenotemark{a}} & 
\colhead{MJD2\tablenotemark{b}} & 
\colhead{$\Delta t\,\tablenotemark{c}\,\, \left[ \rm hrs\right]$} & 
\colhead{$\tau \,\tablenotemark{d}\,\,\left[ \rm hrs\right]$} }
\startdata
52412.15 & 52412.36 &     4.89 &     13.5 $\pm$      0.4\\
52412.36 & 52413.00 &    15.40 &    -19.9 $\pm$     -0.4\\
52413.00 & 52413.14 &     3.49 &      5.9 $\pm$      0.2\\
52414.47 & 52415.14 &    16.06 &    -19.6 $\pm$     -0.3\\
52417.37 & 52417.43 &     1.57 &    -15.6 $\pm$     -1.4\\
52429.30 & 52429.35 &     1.10 &    -19.6 $\pm$     -5.6\\
52429.35 & 52429.61 &     6.09 &    -15.2 $\pm$     -0.8\\
52432.32 & 52432.13 &     4.49 &    -16.5 $\pm$     -1.1\\
52440.11 & 52440.30 &     4.68 &     10.8 $\pm$      0.7\\
52440.30 & 52440.37 &     1.58 &     15.3 $\pm$      4.4\\
52440.37 & 52440.43 &     1.61 &     11.4 $\pm$      2.0\\
52444.12 & 52444.39 &     6.26 &    -17.2 $\pm$     -0.9\\
\enddata
\tablenotetext{a}{\hspace*{0.2cm} Centered MJD of first observation}
\tablenotetext{b}{\hspace*{0.2cm} Centered MJD of second observation}
\tablenotetext{c}{\hspace*{0.2cm} Time difference between observations}
\tablenotetext{d}{\hspace*{0.2cm} $e$-folding time, positive and negative values denotes 
the fastest exponential increase and decrease constants, respectively.}
\end{deluxetable}
%
% HARDENING
%
\begin{deluxetable}{cccc}
\tablecaption{Fastest changes of 3-25 keV photon index. \label{hard}}
\tablewidth{0pt}
\tablehead{
\colhead{MJD1\tablenotemark{a}} & 
\colhead{MJD2\tablenotemark{b}} & 
\colhead{$\Delta t\,\tablenotemark{c}\,\, \left[ \rm hrs\right]$} & 
\colhead{$\Delta \Gamma \,/\,\Delta t\,\tablenotemark{d}\,\,\left[ \rm hrs^{-1}\right]$} }
\startdata
52412.36 & 52413.00 &    15.40 &           0.015 $\pm$           0.001\\
52413.00 & 52413.14 &     3.49 &          -0.083 $\pm$           0.007\\
52431.09 & 52431.34 &     5.98 &          -0.026 $\pm$           0.004\\
52440.43 & 52441.12 &    16.65 &           0.015 $\pm$           0.002\\
52444.12 & 52444.39 &     6.26 &           0.036 $\pm$           0.004\\
\enddata
\tablenotetext{a}{\hspace*{0.2cm} Centered MJD of first observation}
\tablenotetext{b}{\hspace*{0.2cm} Centered MJD of second observation}
\tablenotetext{c}{\hspace*{0.2cm} Time difference between observations in hours}
\tablenotetext{d}{\hspace*{0.2cm} Change in photon index per 1 hr, negative values denote spectral hardening}
\end{deluxetable}
\end{document}